\documentclass[longauth]{aa}

\usepackage{graphicx}
\usepackage{multirow}
\usepackage{txfonts}
\usepackage{hyperref}
\usepackage{lscape}
\usepackage{caption}
\bibpunct{(}{)}{;}{a}{}{,} 

\begin{document} 

   \title{The radio spectral energy distribution of infrared-faint radio
   sources\thanks{Based on observations carried out with the
   IRAM Plateau de Bure Interferometer. IRAM is supported by INSU/CNRS (France), MPG (Germany) and IGN (Spain).}}


   \author{A.~Herzog\inst{1,2,3}
  	  \and
  	  R.~P.~Norris\inst{3}
       \and
       E.~Middelberg\inst{1}
  	  \and
 	  N.~Seymour\inst{4}
 	  \and
  	  L.~R.~Spitler\inst{2,5}
    	  \and
    	  B.~H.~C.~Emonts\inst{6}
       \and
    	  T.~M.~O.~Franzen\inst{4}
    	  \and
    	  R.~Hunstead\inst{7}
    	  \and
    	  H.~T.~Intema\inst{8}
 	  \and
 	  J.~Marvil\inst{3}
 	  \and
 	  Q.~A.~Parker\inst{2,5,9}
    	  \and
    	  S.~K.~Sirothia\inst{10,11,12}
			\and
			N.~Hurley-Walker\inst{4}
    	  \and
    	   M.~Bell\inst{3}
    	   \and
    	   G.~Bernardi\inst{11,12,13}
    	   \and
			J.~D.~Bowman\inst{14} 
			\and
			F.~Briggs\inst{15}
			\and
			R.~J.~Cappallo\inst{16}
			\and
			J.~R.~Callingham\inst{7,17,3}
			\and
			A.~A.~Deshpande\inst{18}
			\and
			K.~S.~Dwarakanath\inst{18}
			\and 
			B.-Q.~For\inst{19}
			\and
			L.~J.~Greenhill\inst{13}
			\and
			P.~Hancock\inst{4,17}
			\and
			B.~J.~Hazelton\inst{20} 
			\and
			L.~Hindson\inst{21}
			\and
			M.~Johnston-Hollitt\inst{21}
			\and
			A.~D.~Kapi\'{n}ska\inst{19,17}
			\and
			D.~L.~Kaplan\inst{22} 
			\and
			E.~Lenc\inst{7,17}
			\and
			C.~J.~Lonsdale\inst{16} 
			\and
			B.~McKinley\inst{23,17}
			\and
			S.~R.~McWhirter\inst{16}
			\and
			D.~A.~Mitchell\inst{3,17}
			\and
			M.~F.~Morales\inst{20}
			\and
			E.~Morgan\inst{24}
			\and
			J.~Morgan\inst{4}
			\and
			D.~Oberoi\inst{25}
			\and
			A.~Offringa\inst{26}
			\and
			S.~M.~Ord\inst{4,17}
			\and
			T.~Prabu\inst{18}
			\and
			P.~Procopio\inst{23}
			\and
			N.~Udaya~Shankar\inst{18} 
			\and
			K.~S.~Srivani\inst{18}
			\and
			L.~Staveley-Smith\inst{19,17}
			\and
			R.~Subrahmanyan\inst{18,17} 
			\and
			S.~J.~Tingay\inst{4,17}
			\and
			R.~B.~Wayth\inst{4,17} 
			\and
			R.~L.~Webster\inst{23,17} 
			\and
			A.~Williams\inst{4}
			\and
			C.~L.~Williams\inst{24}
			\and
			C.~Wu\inst{19}
			\and
			Q.~Zheng\inst{21}
    	  \and
    	  K.~W.~Bannister\inst{3}
    	  \and
    	  A.~P.~Chippendale\inst{3}
    	  \and
    	  L.~Harvey-Smith\inst{3}
    	  \and
    	  I.~Heywood\inst{3}
    	  \and
    	  B.~Indermuehle\inst{3}
    	  \and
    	  A.~Popping\inst{19,17,3}
    	  \and
    	  R.~J.~Sault\inst{3,23}
    	  \and
    	  M.~T.~Whiting\inst{3}
          }

   \institute{Astronomisches Institut, Ruhr-Universit\"at Bochum, Universit\"atsstr. 150, 44801 Bochum, Germany\\
              \email{herzog@astro.rub.de}
         \and
             Macquarie University, Sydney, NSW 2109, Australia
         \and
            CSIRO Astronomy and Space Science, Marsfield, PO Box 76, Epping, NSW
            1710, Australia
          	\and
          	International Centre for Radio Astronomy Research, Curtin University, Bentley, WA 6102, Australia
          \and
             Australian Astronomical Observatory, PO Box 915, North Ryde, NSW
             1670, Australia
          \and
          	Centro de Astrobiolog\'{i}a (INTA-CSIC), Ctra de Torrej\'{o}n a
          	Ajalvir, km 4, 28850 Torrej\'{o}n de Ardoz, Madrid, Spain
          \and
          	Sydney Institute for Astronomy, School of Physics, The University of
          	Sydney, NSW 2006, Australia
          \and
          	National Radio Astronomy Observatory, P.O. Box O, 1003 Lopezville
          	Road, Socorro, NM 87801, USA
          \and
          	Department of Physics, Chong Yeut Ming Physics Building, The
           University of Hong Kong, Pokfulam, Hong Kong
          \and
          	National Centre for Radio Astrophysics, TIFR, Post Bag 3, Pune
          	University Campus, 411007, Pune, India
          	\and
          	SKA SA, 3rd Floor, The Park, Park Road, Pinelands, 7405, South Africa
          	\and
			Department of Physics and Electronics, Rhodes University, PO Box 94,
			Grahamstown, 6140, South Africa
			\and
			Harvard-Smithsonian Center for Astrophysics, 60 Garden Street, Cambridge, MA
			02138, USA
          	\and
          	School of Earth and Space Exploration, Arizona State
          	University, Tempe, AZ 85287, USA
          	\and
          	Research School of Astronomy and Astrophysics, Australian National University, Canberra, ACT 2611, Australia
          	\and
          	MIT Haystack Observatory, Westford, MA 01886, USA
          	\and
          	ARC Centre of Excellence for All-sky Astrophysics (CAASTRO)
          	\and
          	Raman Research Institute, Bangalore 560080, India
            \and
          	International Centre for Radio Astronomy Research, University of
          	Western Australia, 35 Stirling Hwy, Crawley, WA, 6009, Australia
          	\and
          	Department of Physics, University of Washington, Seattle, WA 98195, USA
          	\and
          	School of Chemical \& Physical Sciences, Victoria University
          	of Wellington, PO Box 600, Wellington 6140, New Zealand
          	\and
          	Department of Physics, University of Wisconsin--Milwaukee,
          	Milwaukee, WI 53201, USA
          	\and
          	School of Physics, The University of Melbourne, Parkville, VIC 3010, Australia
           	\and
          	Kavli Institute for Astrophysics and Space Research, Massachusetts Institute of Technology, Cambridge, MA 02139, USA
          	\and
          	National Centre for Radio Astrophysics, Tata Institute for Fundamental Research, Pune 411007, India
          	\and
          	The Netherlands Institute for Radio Astronomy (ASTRON), Postbus 2,
          	7990 AA Dwingeloo, The Netherlands
          	 }

   \date{Received 20 July 2015 / Accepted 08 July 2016}

 
  \abstract
   {Infrared-faint radio sources~(IFRS) are a class of radio-loud~(RL) active
   galactic nuclei~(AGN) at high redshifts~($z\geq 1.7$) that are characterised
   by their relative infrared faintness, resulting in enormous radio-to-infrared
   flux density ratios of up to several thousand.
   }
   {Because of their optical and infrared faintness, it is very
   challenging to study IFRS at these wavelengths. However, IFRS are
   relatively bright in the radio regime with 1.4\,GHz flux densities of a few
   to a few tens of mJy. Therefore, the radio regime is the most promising
   wavelength regime in which to constrain their nature. We aim to test
   the hypothesis that IFRS are young AGN, particularly GHz
   peaked-spectrum~(GPS) and compact steep-spectrum~(CSS) sources that have a
   low frequency turnover.}
   {We use the rich radio data set available for the Australia Telescope Large
   Area Survey fields, covering the frequency range between 150\,MHz and
   34\,GHz with up to 19~wavebands from different telescopes, and build
   radio spectral energy distributions~(SEDs) for 34~IFRS. We then
   study the radio properties of this class of object with respect to turnover, spectral index, and behaviour
   towards higher frequencies. We also present the highest-frequency radio
   observations of an IFRS, observed with the Plateau de Bure Interferometer at
   105\,GHz, and model the multi-wavelength and radio-far-infrared SED of this
   source.}
   {We find IFRS usually follow single power laws down to observed frequencies
   of around 150\,MHz. Mostly, the radio SEDs are steep ($\alpha <
   -0.8$; $74^{+6}_{-9}$\%), but we also find ultra-steep SEDs ($\alpha < -1.3$;
   $6^{+7}_{-2}$\%). In particular, IFRS show statistically significantly
   steeper radio SEDs than the broader RL AGN population. Our analysis reveals
   that the fractions of GPS and CSS sources in the population
   of IFRS are consistent with the fractions in the broader RL AGN population.
   We find that at least $18^{+8}_{-5}$\% of IFRS contain young AGN, although
   the fraction might be significantly higher as suggested by the steep SEDs and the compact
   morphology of IFRS. The detailed multi-wavelength SED modelling of one IFRS
   shows that it is different from ordinary AGN, although it is consistent with a composite
   starburst-AGN model with a star formation rate of 170\,$M_\odot$\,yr$^{-1}$.}
   {}


   \maketitle
%

\section{Introduction}
\label{introduction}

Infrared-faint radio sources~(IFRS) are comparatively bright radio sources with
a faint or absent near-infrared counterpart. They were serendipitously
discovered in the Chandra Deep Field-South~(CDFS) by \citet{Norris2006} in the
Australia Telescope Large Area Survey~(ATLAS) 1.4\,GHz map and the co-located
\textit{Spitzer} Wide-area Infrared Extragalactic Survey~(SWIRE;
\citealp{Lonsdale2003}) infrared~(IR) map. Based on the SEDs of ordinary
galaxies, it was expected that every object in the deep radio survey (rms of
36\,$\mu$Jy\,beam$^{-1}$ at 1.4\,GHz in CDFS) would have a counterpart in the
SWIRE survey~(rms of $\sim 1\,\mu$Jy at 3.6\,$\mu$m).
However, \citeauthor{Norris2006} found 22~radio sources with 1.4\,GHz flux
densities of a few or a few tens of mJy without 3.6\,$\mu$m counterpart and
labelled them as IFRS. Later, IFRS were also found in the European Large Area IR
space observatory Survey South~1~(ELAIS-S1) field, the \textit{Spitzer}
extragalactic First Look Survey~(xFLS) field, the Cosmological Evolution
Survey~(COSMOS) field, the European Large Area IR space observatory Survey
North~1~(ELAIS-N1) field, and the Lockman Hole
field~\citep{Middelberg2008ELAIS-S1,GarnAlexander2008,Zinn2011,Banfield2011,Maini2013submitted},
resulting in around 100~IFRS known in deep fields.\par

While IFRS were originally defined as radio sources without IR counterpart in
the first works, \citet{Zinn2011} set two criteria for the survey-independent
selection of IFRS:
\begin{enumerate}[(i)]
  \item radio-to-IR flux density ratio
    $S_{1.4\,\textrm{GHz}}/S_{3.6\,\mu\textrm{m}} > 500$~, and

  \item 3.6\,$\mu$m flux density $S_{3.6\,\mu\textrm{m}} < 30\,\mu$Jy~.
\end{enumerate}
The first criterion accounts for the enormous radio-to-IR flux density ratios
resulting from the solid radio detection and the IR faintness. These ratios
identify IFRS as clear outliers. The second criterion selects objects at
cosmologically significant redshifts because of cosmic dimming or heavily
obscured objects.\par

\citet{Collier2014} followed a different approach than used in the previous
studies and searched for IFRS based on shallower data, but in a much larger
area. Using the Unified Radio Catalog~(URC; \citealp{Kimball2008}) based on the
NRAO VLA Sky Survey~(NVSS; \citealp{Condon1998}) and IR data from the all-sky
Wide-Field Infrared Survey Explorer~(WISE; \citealp{Wright2010}), they found
1317~IFRS fulfilling both selection criteria from \citet{Zinn2011}. Whereas some
of the IFRS in deep fields are lacking a 3.6\,$\mu$m counterpart, all IFRS from
the catalogue compiled by \citeauthor{Collier2014} have a detected 3.4\,$\mu$m
counterpart. Also, these sources are on average radio-brighter than the IFRS in
deep fields.\par

Since the first IFRS were identified, it has been argued that these objects
might be radio-loud~(RL) active galactic nuclei~(AGN) at high redshifts~($z\gtrsim
1$), potentially heavily obscured by dust~\citep{Norris2006,Norris2011}. Whereas
other explanations like pulsars have been ruled out~\citep{Cameron2011}, the
suggested high redshifts of IFRS have been confirmed by \citet{Collier2014} and
\citet{Herzog2014}; all spectroscopic redshifts are in the range $1.7\lesssim z
\lesssim 3.0$. The first two very long baseline interferometry~(VLBI) detections of IFRS
were carried out by \citet{Norris2007} and \citet{Middelberg2008IFRS_VLBI} who
targeted six IFRS in total and show that at least some IFRS have high
brightness temperatures, indicating the presence of an AGN. Recently,
\citet{Herzog2015a} found compact cores in the majority of IFRS based on a large
sample of 57~sources. \citet{Middelberg2011} show that IFRS have significantly
steeper radio SEDs (median index\footnote{The spectral index~$\alpha$ is defined
as $S\propto \nu^\alpha$ throughout this paper where $S$ is the flux
density and $\nu$ the frequency.} of $-1.4$ between 1.4\,GHz and 2.4\,GHz)
than ordinary AGN.\par

An overlap between the populations of IFRS on the one hand and GHz
peaked-spectrum~(GPS) and compact steep-spectrum~(CSS) sources on the other hand
is suggested and found by \citet{Middelberg2011}, \citet{Collier2014} and
\citet{Herzog2015a}. GPS sources are very compact and powerful AGN with linear
sizes below 1\,kpc, showing a turnover in their radio spectral energy
distribution~(SED) at frequencies of 500\,MHz or higher. CSS sources are
similarly powerful, but are more extended (linear sizes of a few or a few tens
of kpc) and show their turnover at frequencies below 500\,MHz~(e.g.\
\citealp{ODea1998,Randall2011}). Further, CSS sources are characterised by their steep
radio SEDs ($\alpha \lesssim -0.5$). GPS and CSS sources are usually
considered to be young versions of extended radio galaxies, but it has also been
suggested that they are frustrated AGN confined by dense
gas~\citep{ODea1991} or dying radio sources~\citep{Fanti2009b}.\par

Modelling the multi-wavelength SED of IFRS was accomplished by
\citet{GarnAlexander2008}, \citet{Huynh2010}, \citet{Herzog2014}, and
\citet{Herzog2015b}, and shows that these sources can only be modelled as
high-redshift RL AGN, potentially suffering from heavy dust extinction. The
strong link between IFRS and high-redshift radio galaxies~(HzRGs)---first
suggested by \citeauthor{Huynh2010} and \citet{Middelberg2011} and later
emphasised by \citet{Norris2011}---has also been found in the modelling by
\citet{Herzog2015b}. HzRGs are massive galaxies at high redshifts ($1\leq
z \leq 5.2$) which are expected to be the progenitors of the most massive
elliptical galaxies in the local universe (e.g.\ \citealp{Seymour2007,deBreuck2010}). They host AGN
and undergo heavy star forming activity. IFRS have a significantly higher sky
density than HzRGs (a few IFRS per square degree versus around 100~HzRGs known
on the entire sky) and are suggested to be higher-redshift or less luminuous
siblings of these massive galaxies.\par

The correlation between $K$~band magnitude and redshift has been known for radio
galaxies~(e.g.\ \citealp{Lilly1984,Willott2003,Rocca-Volmerange2004}) for three
decades and was used to find radio galaxies at high redshifts. In particular,
HzRGs were also found to follow this correlation~\citep{Seymour2007}. Although
IFRS are on average fainter than HzRGs in the near-IR regime, an overlap between
both samples exists. \citet{Norris2011} suggest that IFRS might
also follow a correlation between near-IR flux density and redshift. This
suggestion has been supported by \citet{Collier2014} and \citet{Herzog2014} who
find that those IFRS with spectroscopic redshifts are consistent with this
suggested correlation. Similarly, ultra-steep radio spectra~($\alpha \lesssim -1.0$) are
known to be successful tracers of high-redshift galaxies~(e.g.\
\citealp{Tielens1979,McCarthy1991,Roettgering1994}). The classes of HzRGs and
IFRS were both found to have steep radio spectra~\citep{Middelberg2011}.\par

Studying IFRS in the optical and IR regime is challenging because of their
faintness at these frequencies. In contrast, IFRS are relatively bright in the
radio regime, making detailed radio studies feasible. Since the radio emission
of RL galaxies is dominated by the AGN, radio studies of IFRS can provide
insights into the characteristics of the active nucleus, e.g.\ its age.\par

This paper aims at studying the broad radio SEDs of IFRS, spanning a frequency
range of more than two orders of magnitude. In Sect.~\ref{data}, we present our
sample of 34~IFRS from the ATLAS fields and describe the available data for the
ELAIS-S1 and CDFS fields which includes the first data on IFRS below 610\,MHz
and above 8.6\,GHz. Among others, we are using data of two of the new-generation
radio telescopes and Square Kilometre Array~(SKA; \citealp{Dewdney2009})
precursors, Murchison Widefield Array~(MWA; \citealp{Lonsdale2009,Tingay2013})
and Australian Square Kilometre Array Pathfinder~(ASKAP;
\citealp{Johnston2007,Johnston2008,DeBoer2009}). We also describe the Plateau de
Bure Interferometer~(PdBI) observations---the highest-frequency radio
observations of an IFRS so far---and ancillary data of one IFRS in the xFLS
field. Based on the available data, we build and fit radio SEDs for the IFRS in
the ATLAS fields in Sect.~\ref{building_fitting_radioSEDs} and analyse them with
respect to spectral index, turnover, and high-frequency behaviour in
Sect.~\ref{discussion_radioSEDs}. In Sect.~\ref{IFRS_xFLS478}, we present a
multi-wavelength and radio SED modelling for the IFRS observed with the
PdBI. Our results are summarised in Sect.~\ref{conclusion}. The photometric data
obtained in Sect.~\ref{data} are summarised in Appendix~\ref{datasection}.
Throughout this paper, we use flat $\Lambda$CDM cosmological parameters
$\Omega_\Lambda = 0.7$, $\Omega_\textrm{M} = 0.3$, $H_0 =
70$~km~s$^{-1}$~Mpc$^{-1}$, and the calculator by \citet{Wright2006}. The linear
scale in $\Lambda$CDM cosmology is limited in the redshift range $0.5\leq z \leq
12$ between 4\,kpc/arcsec and 8.5\,kpc/arcsec. Following
\citet{Cameron2011_beta}, we calculate $1\sigma$ confidence intervals of
binomial population proportions based on the Bayesian approach.\par

\section{Observations and data}
\label{data}

Aiming at building the broad radio SEDs of a larger number of IFRS, we based our
sample on the IFRS catalogue compiled by \citet{Zinn2011}. This catalogue
contains 55~IFRS in the ELAIS-S1, CDFS, xFLS and COSMOS fields. Because of the
rich radio data set in the ELAIS-S1 and CDFS fields, we limited our study to
IFRS in these two fields. However, we discarded the source~ES11 from our
sample since it was recently found to be putatively associated with a
3.6\,$\mu$m SWIRE source in high-resolution radio observations~(Collier et al.,
in prep.), not fulfilling the selection criteria from \citeauthor{Zinn2011} any
more. Thus, we used 28~IFRS from the sample presented by
\citeauthor{Zinn2011} for our study: 14~IFRS in ELAIS-S1, and 14 in CDFS.\par

\citet{Maini2013submitted} presented a catalogue of IFRS based on the
deeper \textit{Spitzer} Extragalactic Representative Volume Survey~(SERVS) near-
and mid-IR~data, also covering parts of the ELAIS-S1 and the CDFS fields.
Because of the deeper 3.6\,$\mu$m data, \citeauthor{Maini2013submitted} were
able to identify some IFRS that were not listed in the IFRS catalogue from
\citet{Zinn2011}. These sources were undetected in the shallower SWIRE survey.
However, because of their 1.4\,GHz flux densities of around 1\,mJy, they did not
fulfil criterion~(i) from \citeauthor{Zinn2011} but meet the criterion based on
a SERVS detection below the SWIRE limit. In order to study the less extreme
versions of IFRS, \citeauthor{Maini2013submitted} lowered the first IFRS
selection criterion from \citeauthor{Zinn2011} and included sources with a
radio-to-IR flux density ratio above 200 in their sample. Aiming at studying the
originally very extreme class of IFRS, in our work, we limited our sample to a
radio-to-IR flux density ratio of 500 for the definition of IFRS and added only
those sources in ELAIS-S1 and CDFS from \citeauthor{Maini2013submitted} to our
sample that fulfil this stronger criterion. Adding one IFRS in ELAIS-S1 and five
IFRS in CDFS, we ended up with a sample size of 34~IFRS for our radio SED study:
15 in ELAIS-S1 and 19 in CDFS. Throughout this paper, we use identifiers from
\citeauthor{Zinn2011} and \citeauthor{Maini2013submitted} which are identical to
the identifiers in the first ATLAS data release~(DR1) presented by
\citet{Norris2006} and \citet{Middelberg2008ELAIS-S1}.\par

We describe our radio data in Sects.~\ref{data_ELAIS} and \ref{data_CDFS} for
ELAIS-S1 and CDFS, respectively. All observations are summarised in
Tables~\ref{tab:observations_ELAIS} and \ref{tab:observations_CDFS}, listing
frequency, telescope, angular resolution, maximum sensitivity, and the number
of detected IFRS, undetected IFRS, and IFRS outside the field, respectively.
All photometric data are listed in Appendix~\ref{datasection} in
Tables~\ref{tab:data_ELAIS} and \ref{tab:data_CDFS} for ELAIS-S1 and CDFS,
respectively. We comment on our cross-matching approach in
Sect.~\ref{crossmatching} and clarify our way of dealing with flux density
uncertainties in Sect.~\ref{errors}. Issues arising from different angular
resolutions are discussed in Sect.~\ref{resolutioneffects} and a control sample
is introduced in Sect.~\ref{controlsample}. In Sect.~\ref{data_xFLS478}, we
present observations of the IFRS xFLS\,478 with the PdBI, describe the data
calibration, and collect ancillary data.\par
\begin{table*}
	\caption{Characteristics of the observations of the ELAIS-S1 field, covering 15~IFRS from our sample.}
	\label{tab:observations_ELAIS}
 \centering
 \begin{tabular}{c c c c c c c c}
 	\hline \hline
	Frequency        & Telescope & Angular resolution   & Sensitivity & \# det & \# undet & \# outside & Reference \\
	$[\mathrm{MHz}]$ &           & [arcsec$^2$]         & [$\mu$Jy\,beam$^{-1}$]   &        &          &            &           \\ \hline
        200              & MWA       & $140 \times 125$     & 7000        & 13     & 2        & \ldots     & (1)       \\
	610              & GMRT      & $11 \times 11$       & 100         & 14     & \ldots   &	1          & (2)       \\
	843              & MOST      & $62 \times 43$       & 600         & 15     & \ldots   & \ldots     & (3)       \\
	1400             & ATCA      & $12 \times 8$        & 17          & 15     & \ldots   & \ldots     & (4)       \\
	2300             & ATCA      & $33.56 \times 19.90$ & 70          & 15     & \ldots   & \ldots     & (5)       \\
	4800             & ATCA      & $4.6 \times 1.7$     & 130         & 5      & 1        & 9          & (6)       \\
	8640             & ATCA      & $4.6 \times 1.7$     & 130         & 3      & 3        & 9          & (6)       \\
	34\,000          & ATCA      & $7 \times 7$         & 110         & 2      & 1        & 12         & (7)       \\
	\hline
	\end{tabular}
        \tablefoot{Columns 5, 6, and 7 list the number of detected IFRS, undetected IFRS, and IFRS outside the field, respectively.}
        \tablebib{(1)~Hurley-Walker et al. (submitted); (2)~Based on the map
        from Intema et al. (in prep.); (3)~Based on the map from \citet{Randall2012}; (4)~\citet{Franzen2015}; (5)~Based on the map from \citet{Zinn2012}; (6)~\citet{Middelberg2011}; (7)~Emonts et al. (in prep.)}
\end{table*}

\begin{table*}
	\caption{Characteristics of the observations of the CDFS field, covering 19~IFRS from our sample.}
	\label{tab:observations_CDFS}
 \centering
 \begin{tabular}{c c c c c c c c}
 	\hline \hline
	Frequency        & Telescope  & Angular resolution   & Sensitivity           & \# det            & \# undet & \# outside & Reference \\
	$[\mathrm{MHz}]$ &            & [arcsec$^2$]         & [$\mu$Jy\,beam$^{-1}$]  &                   &          &            &           \\ \hline
        150              & GMRT       & $25.2 \times 14.7$   & 2000                  & 12                & 7        & \ldots     & (1)       \\
        200              & MWA        & $135 \times 125$     & 5900                  & 10                & 9        & \ldots     & (2)       \\
        325              & GMRT       & $11 \times 7 $       & 100                   & 18                & 1        & \ldots     & (1)       \\
	610              & GMRT       & $7.7 \times 3.7$     & 100                   & 10                & \ldots   & 9          & (3)       \\
	843              & MOST       & $95 \times 43$       & 1700                  & 8                 & 9        & 2          &          \\
        844              & ASKAP-BETA & $91 \times 56$       & 450                   & 18                & 1        & \ldots     & (4)       \\
	1400             & ATCA       & $16 \times 7$        & 14                    & 19                & \ldots   & \ldots     & (5)       \\
	2300             & ATCA       & $57.15 \times 22.68$ & 70                    & 19                & \ldots   & \ldots     & (6)       \\
	4800             & ATCA       & $4.6 \times 1.7$     & 100                   & 5                 & 3        & 11         & (7)       \\
        5500             & ATCA       & $4.9 \times 2.0$     & 12                    & 2                 & \ldots   & 17         & (8)       \\
	8640             & ATCA       & $4.6 \times 1.7$     & 90                    & 5                 & 3        & 11         & (7)       \\
        20\,000          & ATCA       & $29.1 \times 21.9$   & 40                    & 2\tablefootmark{a}& 10       & 7          & (9)      \\
	34\,000          & ATCA       & $8.2 \times 5.1$     & 30                    & 3                 & \ldots   & 16         & (10)      \\
	\hline
	\end{tabular}
	\tablefoot{Columns 5, 6, and 7 list the number of detected IFRS, undetected IFRS, and IFRS outside the field, respectively.
        \tablefoottext{a}{These two IFRS were also detected in the follow-up observations by \citet{Franzen2014} at 5.5\,GHz, 9\,GHz, and 18\,GHz.}}
        \tablebib{(1)~Based on the map from Sirothia et al. (in prep.);
        (2)~Hurley-Walker et al. (submitted); (3)~Based on the map from Intema
        et al.
        (in prep.); (4)~Based on the map from Marvil et al. (in prep.); (5)~\citet{Franzen2015}; (6)~Based on the map from \citet{Zinn2012}; (7)~\citet{Middelberg2011}; (8)~\citet{Huynh2012}; (9)~\citet{Franzen2014}; (10)~Emonts et al. (in prep.)}
\end{table*}

\subsection{Radio data for ELAIS-S1}
\label{data_ELAIS}

\subsubsection{1.4\,GHz ATLAS DR3 data}
\label{data_ELAIS_1p4GHz}
Since the definition of IFRS is based on the observed 1.4\,GHz flux density, all
IFRS are detected at this frequency. \citet{Zinn2011} used data from ATLAS
DR1~\citep{Norris2006,Middelberg2008ELAIS-S1} for their IFRS catalogue. Here, we
used the recent ATLAS data release~3 (DR3; \citealp{Franzen2015}).
ATLAS DR3 has a resolution of $12\times 8$\,arcsec$^2$ and a sensitivity of $\sim
17\,\mu$Jy\,beam$^{-1}$ (up to $100\,\mu$Jy\,beam$^{-1}$ at the edges) at
1.4\,GHz in ELAIS-S1. \citeauthor{Franzen2015} applied three criteria
for their component catalogue: (1) local rms noise below
100\,$\mu$Jy\,beam$^{-1}$, (2) sensitivity loss arising from bandwidth smearing
below 20\%, and (3) primary beam response at least 40\% of the peak response.
Sources in ATLAS DR3 have been fitted with one or more Gaussians, where each
Gaussian is referred to as a single ``component''. Thus, a source can consist of
one or more components.\par

We extracted all components from the ATLAS DR3 component catalogue by
\citet{Franzen2015} that we deemed to be associated with our 15~IFRS in
ELAIS-S1. Eleven~component counterparts were found for eight~IFRS, fulfilling
all three selection criteria from \citeauthor{Franzen2015} Seven IFRS
did not provide counterparts in ATLAS DR3. These sources are located close to the field
edges and the respective sources in the DR3 map do not fulfil the primary
beam response criterion~(3). Therefore, these components are not listed in the
component catalogue presented by \citeauthor{Franzen2015}
\citet{Middelberg2008ELAIS-S1} used different component selection criteria which
allowed sources at the field edges to be included in their catalogue.\par

Component extraction was performed on those seven IFRS without counterpart in
the DR3 catalogue in the same way as presented by \citet{Franzen2015},
however at the cost of lower beam response and higher local rms noise. Thus,
nine component counterparts were found for the seven remaining IFRS, i.e.\
1.4\,GHz ATLAS DR3 counterparts could be extracted for all IFRS from our sample
in ELAIS-S1.\par

We visually inspected the 1.4\,GHz map along with the 3.6\,$\mu$m SWIRE map to
check whether all components found in our manual cross-matching were associated
with the IFRS. If a source is composed of more than one Gaussian component in
DR3, an these components are clearly separated, and we found a 3.6\,$\mu$m
counterpart for more than one of these radio components, we disregarded those
additional radio components with IR counterparts. Because of their IR
counterparts, these components are probably not radio jets of a spatially
separated galaxy. In this approach, we discarded one out of 20~Gaussian
components found for our 15~IFRS in ELAIS-S1. Therefore, the grouping of
Gaussian components to sources differed from the automatic approach used by
\citet{Franzen2015} in some cases.\par

We extracted integrated flux densities at 1.4\,GHz from ATLAS DR3. If the
counterpart of an IFRS was confirmed to be composite of more than one component
in DR3 as described above, we added the integrated flux densities of the
individual components and propagated the errors. Because of discarding
components as described above, the 1.4\,GHz flux densities of the IFRS in our
sample might differ from the respective numbers in the ATLAS DR3 source
catalogue.\par

The catalogue presented by \citet{Franzen2015} provides a
spectral index~$\alpha^{1.71}_{1.40}$ between 1.40\,GHz and 1.71\,GHz. We list
this information in Table~\ref{tab:data_ELAIS} and used it in our analysis.
However, for three IFRS in ELAIS-S1 located very close to the ATLAS field
edges, the spectral index was not available since these sources were outside
the mosaic field in the higher-frequency subband.\par

\subsubsection{610\,MHz GMRT data}
\label{data_ELAIS_GMRT}
The ELAIS-S1 field was observed with the Giant Metrewave Radio Telescope~(GMRT)
at 610\,MHz with a resolution of $11\times 11$\,arcsec$^2$~(Intema et al., in
prep.) down to a median rms of 100\,$\mu$Jy\,beam$^{-1}$ over large parts of the
field and up to 450\,$\mu$Jy\,beam$^{-1}$ at the edges. Nine out of our 15~IFRS
in ELAIS-S1 were located in the final map of this project. Five more IFRS were
also covered by these observations, but are located outside the final map
where the primary beam response is low and the beam shape is poorly known,
resulting in higher noise and uncertainty. We measured integrated flux densities
from the extended map using \texttt{JMFIT}\footnote{\texttt{JMFIT} is a task of
the Astronomical Image and Processing
System~(AIPS); \url{http://www.aips.nrao.edu/}}---also including data with low
beam response---for all 14~IFRS covered in these observations and accounted for the
higher uncertainty as described in Sect.~\ref{errors}. IFRS~ES1259 was not
targeted by these observations.\par

\subsubsection{200\,MHz GLEAM data}
\label{data_ELAIS_GLEAM}
The Galactic and Extragalactic MWA Survey~(GLEAM) targeted the entire sky south
of $+30^\circ$ declination at $72- 231$\,MHz~\citep{Wayth2015} with the MWA.
Here, we used the GLEAM data release~1~(Hurley-Walker et al.,
submitted). The GLEAM catalogue is based on a deep image, covering the frequency
range between 170\,MHz and 230\,MHz. Each source detected in this deep image was then
re-measured in each of the twenty 8\,MHz-wide subbands between 72\,MHz and
231\,MHz. The beam size in ELAIS-S1 is around $140\times 125$\,arcsec$^2$ and
the rms around 7.0\,mJy\,beam$^{-1}$ in the deep 60\,MHz image. We found
counterparts for 13 out of 15~IFRS in ELAIS-S1. For the two IFRS undetected
in the GLEAM survey, we set conservative $4\sigma$ flux density upper
limits based on the local rms.\par

Since the IFRS counterparts are comparatively faint at these frequencies
and close to the GLEAM detection limit, the uncertainties in the individual
GLEAM subbands are relatively large with respect to the measured flux densities.
Therefore, we used a weighted average of four subbands at a time to increase the
signal-to-noise ratio. Thus, we obtained five flux density data points from the
20~GLEAM subbands for all IFRS detected in the deep image.\par

\subsubsection{843\,MHz MOST data}
\label{data_ELAIS_843MHz}
\citet{Randall2012} presented observations of the ELAIS-S1 field at 843\,MHz
with the Molonglo Observatory Synthesis Telescope~(MOST). The data have a
resolution of $62\times 43$\,arcsec$^2$ and an rms of around
0.6\,mJy\,beam$^{-1}$. The observations from \citeauthor{Randall2012} use the
same frequency and resolution as the Sydney University Molonglo Sky Survey~(SUMSS;
\citealp{Bock1999,Mauch2003}), but are twice as sensitive.\par

To be consistent with MOST observations of the CDFS described below, we measured
flux densities in the same way in both fields using \texttt{JMFIT}. As reported
by \citet{Randall2012}, there are two types of artefacts in their final map: grating rings and radial
spokes, where the former one is relevant for our flux measurements. One of these rings
interferes with one of our sources~(ES1259) and neither a flux density nor an
upper limit could be reliably measured. In SUMSS, this source is also affected
by this artefact.\par

Furthermore, sources in the final map from \citet{Randall2012} are surrounded by
a ring of negative pixel values (``holes''). We accounted for this issue by
fitting a background level and subtracting this background from the measured
flux densities using \texttt{JMFIT}. We found 843\,MHz flux densities for all
15~sources but ES1259. These flux densities were found to be in agreement with
those reported by \citeauthor{Randall2012} and also in agreement with the SUMSS
flux densities for sources listed in that survey catalogue.\par

\subsubsection{2.3\,GHz ATLAS data}
\label{data_ELAIS_2p3GHz}
The ELAIS-S1 field was observed with the Australia Telescope Compact
Array~(ATCA) at 2.3\,GHz~\citep{Zinn2012} as part of the ATLAS survey. The
observations resulted in an rms of 70\,$\mu$Jy\,beam$^{-1}$ and a resolution of
$33.56\times 19.90$\,arcsec$^2$. We cross-matched our IFRS sample with the
source catalogue from \citeauthor{Zinn2012} and found six out of 15~IFRS in
ELAIS-S1 to have a counterpart at 2.3\,GHz.\par

At the positions of all nine IFRS without catalogued 2.3\,GHz counterparts by
\citet{Zinn2012}, unambiguous detections are visible in the 2.3\,GHz map. Seven
of these IFRS are located close to the edges of the field and their 2.3\,GHz
counterparts might therefore not be listed by \citeauthor{Zinn2012} It is
unclear why ES419 and ES427 in the centre of the field do not have catalogued
2.3\,GHz counterparts.\par

Because of these missing 2.3\,GHz counterparts, we measured flux
densities from the 2.3\,GHz map from \citet{Zinn2012} using \texttt{JMFIT} for all IFRS in
ELAIS-S1. For the IFRS with 2.3\,GHz counterparts listed by
\citeauthor{Zinn2012}, we found that the flux densities measured in our work
are in agreement with the flux densities from \citeauthor{Zinn2012} For
consistency, we used 2.3\,GHz flux densities measured in our work for
all 15~IFRS in ELAIS-S1.\par

\subsubsection{Higher-frequency radio data}
\label{data_ELAIS_other}
\citet{Middelberg2011} studied the higher-frequency radio SEDs of IFRS and
observed nine sources in the ELAIS-S1 field with the ATCA at 4.8\,GHz and
8.6\,GHz down to an rms of around 130\,$\mu$Jy\,beam$^{-1}$. Six IFRS from our sample
in ELAIS-S1 were observed in this study, resulting in five detections at 4.8\,GHz
and three detections at 8.6\,GHz. The observations had an angular resolution of
$4.6\times 1.7$\,arcsec$^2$ at both frequencies. We used the integrated flux
densities and flux density upper limits presented by \citeauthor{Middelberg2011}
in our study.\par

Three IFRS from our sample in ELAIS-S1 were observed with the ATCA at 34\,GHz,
resulting in a resolution of around 7\,arcsec and an rms of around
110\,$\mu$Jy\,beam$^{-1}$ (Emonts et al., in prep.). Two of the targeted IFRS
were detected and one IFRS was found to be undetected. We used the related flux
densities and upper limits in our study.\par

\subsection{Radio data for CDFS}
\label{data_CDFS}

\subsubsection{1.4\,GHz ATLAS DR3 data}
\label{data_CDFS_1p4GHz}
The 1.4\,GHz ATLAS DR3 data~\citep{Franzen2015} of the CDFS field with
a resolution of $16\times 7$\,arcsec$^2$ and a sensitivity of $\sim
14\,\mu$Jy\,beam$^{-1}$ (up to $100\,\mu$Jy\,beam$^{-1}$ at the field edges) was
used. We extracted all components from the ATLAS DR3 component catalogue that we
deemed to be associated with our 19~IFRS in CDFS as described in
Sect.~\ref{data_ELAIS_1p4GHz} for the ELAIS-S1 field. We found 29~component
counterparts for 17~IFRS. Counterparts in DR3 for the other two IFRS were
missing because of the primary beam criterion as mentioned in
Sect.~\ref{data_ELAIS_1p4GHz}. Again, component extraction was performed on the
ATLAS DR3 map at the respective positions in the same way as presented by
\citeauthor{Franzen2015}. Three component counterparts were found
for these two IFRS. The resulting component catalogue was analysed and used as
described in Sect.~\ref{data_ELAIS_1p4GHz}. In the visual inspection, we
discarded six Gaussian components.\par

We emphasise that IFRS CS618 is peculiar and differs from all other IFRS in our
sample because of its morphology. In the 1.4\,GHz ATLAS map, this source appears
as a typical double-lobed radio galaxy, consisting of three clearly separated
emission regions, which were fitted by four Gaussian components in DR3. In
Sect.~\ref{S618}, we discuss the characteristics of this source in detail.\par

In-band spectral indices~$\alpha_{1.40}^{1.71}$ between 1.40\,GHz and
1.71\,GHz were also taken from the ATLAS DR3 catalogue~\citep{Franzen2015}
and used in our analysis. They are listed in Table~\ref{tab:data_CDFS}. Because
of the peculiar characteristics of IFRS CS618 discussed in more detail below, no
in-band spectral index was available for this source.\par

\subsubsection{150\,MHz, 325\,MHz, and 610\,MHz GMRT data}
\label{data_CDFS_GMRT}
The maps of the CDFS at 150\,MHz and 325\,MHz~(Sirothia et al., in prep.) are
based on data from the GMRT and have resolutions of $25\times 15$\,arcsec$^2$
and $11\times 7$\,arcsec$^2$, respectively. The sensitivities reach around
2\,mJy\,beam$^{-1}$ and 100\,$\mu$Jy\,beam$^{-1}$, respectively. We found
counterparts for twelve IFRS at 150\,MHz and measured their flux densities using
\texttt{JMFIT}. Seven IFRS remained undetected at 150\,MHz. At 325\,MHz, we
found counterparts for 18~IFRS using \texttt{JMFIT}. The only undetected IFRS at
this frequency is CS94. This source is located in an area where the noise is significantly higher and neither a
counterpart nor a flux density upper limit could be reliably determined for this
IFRS.\par

The TIFR GMRT Sky Survey\footnote{\url{http://tgss.ncra.tifr.res.in/}}~(TGSS)
aims to observe 37\,000\,deg$^2$ at 150\,MHz with a sensitivity of
7\,mJy\,beam$^{-1}$. TGSS DR5 (November 2012) covers
parts of the CDFS at a sensitivity of around 8\,mJy\,beam$^{-1}$, and is assumed
to have an uncertainty of 25\% in flux density. Three IFRS from our sample are
detected in TGSS DR5 and we found our flux densities measured with
\texttt{JMFIT} in agreement with the TGSS results. However, for consistency, we
used our flux densities for all sources in our study at 150\,MHz and
325\,MHz.\par

Three parts of the CDFS were observed with one pointing each with the GMRT at
610\,MHz~(Intema et al., in prep.). These pointings were centred on the IFRS
CS114, CS194, and CS703. Five additional IFRS (CS97, CS265, CS292, CS618, CS713)
are also located in the pointing fields. These observations reach
sensitivities of 95\,$\mu$Jy\,beam$^{-1}$, 150\,$\mu$Jy\,beam$^{-1}$, and
80\,$\mu$Jy\,beam$^{-1}$, respectively, at a resolution of around $7.7\times
3.7$\,arcsec$^2$. We measured flux densities from the maps using \texttt{JMFIT}
and found 610\,MHz counterparts for all eight~IFRS.\par

\subsubsection{200\,MHz GLEAM data}
\label{data_CDFS_GLEAM}
We cross-matched our IFRS sample in CDFS with the final GLEAM
catalogue~(Hurley-Walker et al., submitted), selected at 200\,MHz with
60\,MHz bandwidth as presented in Sect.~\ref{data_ELAIS_GLEAM}. For each
detected source, the catalogue provides flux densities in 20~subbands between
72\,MHz and 231\,MHz, each with a bandwidth of 8\,MHz. The catalogue has an
angular resolution of around $135\times 125$\,arcsec$^2$ and an average rms of
around 5.9\,mJy\,beam$^{-1}$ in the deep 60\,MHz image in CDFS. We found
counterparts for ten out of 19~IFRS in CDFS. For the nine IFRS undetected
in the GLEAM survey, we set conservative $4\sigma$ flux density upper limits
based on the local rms. We averaged four GLEAM subbands at a time as
described in Sect.~\ref{data_ELAIS_GLEAM}.\par

\subsubsection{843\,MHz MOST data}
\label{data_CDFS_MOST}
The CDFS was observed with MOST at 843\,MHz over several epochs in 2008, very
similar to the observations of the ELAIS-S1 field described in
Sect.~\ref{data_ELAIS_843MHz}. In CDFS, the map reaches a sensitivity of around
1.7\,mJy\,beam$^{-1}$ at a resolution of $95\times 43$\,arcsec$^2$. We measured
flux densities in the same way as described for ELAIS-S1 in
Sect.~\ref{data_ELAIS_843MHz}. Two IFRS are located outside the field and one
IFRS is affected by radial spokes. Of the remaining 16~IFRS from our sample,
nine sources provided a counterpart at 843\,MHz; all other sources were
undetected.\par

\subsubsection{844\,MHz ASKAP-BETA data}
\label{data_CDFS_BETA}
The six antennas of the Boolardy Engineering Test Array~(BETA;
\citealp{Hotan2014}), a subset of ASKAP\footnote{See
\url{http://www.atnf.csiro.au/projects/askap}}, were used by the ASKAP
Commissioning and Early Science~(ACES) team to observe a region of around
22\,deg$^2$ at 844\,MHz~(Marvil et al., in prep.). The rms in this field is
around 450\,$\mu$Jy\,beam$^{-1}$ and the angular resolution $91\times
56$\,arcsec$^2$. The field includes the CDFS, i.e.\ all IFRS in CDFS were
covered by these observations and we found counterparts for 18~sources using
\texttt{JMFIT}. Flux densities for sources detected both in the MOST
observations~(Sect.~\ref{data_CDFS_MOST}) and in the ASKAP-BETA observations
agree within the uncertainties.\par

\subsubsection{2.3\,GHz ATLAS data}
\label{data_CDFS_2p3GHz}
The 2.3\,GHz survey of the CDFS presented by \citet{Zinn2012} has an rms of
70\,$\mu$Jy\,beam$^{-1}$ at a resolution of $57.15\times 22.68$\,arcsec$^2$. 13
out of 19~IFRS in the CDFS field have a 2.3\,GHz counterpart listed in the
source catalogue from \citeauthor{Zinn2012} The other six IFRS show 2.3\,GHz
counterparts in the map, too. Four sources are located close to the field edges
and therefore might not be listed in the 2.3\,GHz source catalogue. CS265 and
CS538 are in the centre of the field and it is unclear why their 2.3\,GHz
counterparts are not listed in the source catalogue from
\citeauthor{Zinn2012}\par

To obtain 2.3\,GHz flux densities for all IFRS in CDFS, we measured flux
densities of all IFRS as described in Sect.~\ref{data_ELAIS_2p3GHz}. For the
13~IFRS with 2.3\,GHz counterpart presented by \citet{Zinn2012}, we found the
2.3\,GHz flux densities measured in our work to be consistent with the flux
densities listed by \citeauthor{Zinn2012} For consistency in our study, we used
our own 2.3\,GHz flux densities for all IFRS in CDFS.\par

\subsubsection{Higher-frequency radio data}
\label{data_CDFS_other}

\citet{Middelberg2011} observed eight IFRS from our sample in CDFS with the ATCA
at 4.8\,GHz and 8.6\,GHz at a resolution of $4.6\times 1.7$\,arcsec$^2$ and an
rms of around 90\,$\mu$Jy\,beam$^{-1}$ and 100\,$\mu$Jy\,beam$^{-1}$, respectively. Five of
these IFRS were detected both at 4.8\,GHz and 8.6\,GHz, the other three IFRS
remained undetected at both frequencies. We used the integrated flux densities
from these observations in our study.\par

\citet{Huynh2012} observed the 0.25\,deg$^2$ field of the extended CDFS~(eCDFS)
with the ATCA at 5.5\,GHz at a resolution of $4.9\times 2.0$\,arcsec$^2$, resulting in an rms
of 12\,$\mu$Jy\,beam$^{-1}$. Two of our IFRS---CS520 and CS415---lie in the
field covered by this survey and both were detected.
We extracted integrated flux densities with respective errors from
\citeauthor{Huynh2012}\par

Higher-frequency data used for our study were taken from the Australia Telescope
20\,GHz~(AT20G) deep pilot survey~\citep{Franzen2014}. Among other fields,
this survey targeted the CDFS at 20\,GHz at resolution of $29.1\times 21.9$\,arcsec$^2$ down
to an rms of 0.3\,mJy\,beam$^{-1}$ or 0.4\,mJy\,beam$^{-1}$. Two~IFRS---CS265
and CS603---were detected, whereas ten IFRS remained undetected at 20\,GHz at
this sensitivity and the other seven IFRS were located outside the final AT20G
field.\par

This project also included follow-up observations at 18\,GHz, 9\,GHz, and
5.5\,GHz of the sources detected at 20\,GHz. The angular resolutions were around
10\,arcsec, 25\,arcsec, and 40\,arcsec at 18\,GHz, 9\,GHz, and 5.5\,GHz,
respectively. The IFRS CS265 and CS603 were both detected at all three follow-up
frequencies. We used the integrated flux densities at all four frequencies from
the AT20G project~\citep{Franzen2014} for CS265 and CS603 and conservative flux
density upper limits at 20\,GHz for the undetected IFRS in the survey field.\par

Three IFRS from our sample (CS114, CS194, CS703) were observed with the ATCA at
34\,GHz, resulting in a resolution of $8.2\times 5.1$\,arcsec$^2$ and an rms of
around 30\,$\mu$Jy\,beam$^{-1}$ (Emonts et al., in prep.). All three targeted
IFRS were detected.\par

\subsection{Cross-matching of radio data}
\label{crossmatching}
Cross-matching of data from different catalogues---characterised by different
angular resolution, sensitivity, and observing frequency---is a crucial step in
order to gain broad-band information about the SEDs of astrophysical objects.
Sophisticated methods such as the likelihood ratio~\citep{Sutherland1992} or
Bayesian approaches~\citep{Fan2015arXiv} were unnecessary in our case as we were
matching radio data with other radio data, the sky density of objects in these different
surveys is comparatively low, and the mean distance between sources is much
greater than our beamwidth. Thus, when cross-matching different catalogues, we
followed a nearest-neighbour approach and checked by eye whether the
cross-matching was correct and unambiguous.\par

\subsection{Flux density uncertainties}
\label{errors}
Uncertainties on flux densities of radio sources are composed of a number of
different contributions, namely errors on gain factors and source fitting, the local
background rms noise, \texttt{CLEAN}ing errors and other errors. Since this work
is based on radio data from several projects, a proper derivation of errors for
individual flux density measurements is challenging due to the different
characteristics of telescopes, surveys, and observations.\par

For flux densities~$S$ measured in this work, we derived the related flux
density uncertainties~$S_\mathrm{err}$ using the approach
\begin{equation}
S_\mathrm{err} = \sqrt{(a_\mathrm{calib}\cdot S)^2 + dS^2 +
(a_\mathrm{edge}\cdot S)^2}~,
\end{equation}
where $a_\mathrm{calib}$ is the fractional calibration error, $dS$ is the flux
density error obtained from the source fitting using \texttt{JMFIT},
and $a_\mathrm{edge}$ is an additional fractional error for some
observations that applies when a source is located close to the primary beam
edges. We note that the error obtained from \texttt{JMFIT} includes exclusively
the rms of the image since the error resulting from fitting a Gaussian to the
source is tiny and is therefore neglected in this task.\par

For observations with the GMRT (150\,MHz, 325\,MHz, 610\,MHz), we assumed a
calibration uncertainty of 25\%, i.e.\ $a_\mathrm{calib} = 0.25$. The quoted
accuracy of the 843\,MHz flux densities from MOST in ELAIS-S1 is
0.05~\citep{Randall2012}. Since the MOST observations of the CDFS were carried
out and calibrated in the same way, we also used an accuracy of 0.05. For the
ASKAP-BETA data at 844\,MHz, we set $a_\mathrm{calib} = 0.1$. At 2.3\,GHz, we
assumed an accuracy of 0.1. An additional error applies in the GMRT observations at
610\,MHz for some sources located at the edges of the respective fields because
of pointing errors. In this case, we set $a_\mathrm{edge}$ to 0.15. In all other
cases, $a_\mathrm{edge}$ was set to zero.\par

When using data from published catalogues (1.4\,GHz, 4.8\,GHz, 5.5\,GHz,
8.6\,GHz, 9\,GHz, 18\,GHz, 20\,GHz), we used the flux density errors quoted in
the respective catalogue. For GLEAM counterparts, we added in quadrature a
fractional uncertainty of 0.08 to the catalogued
source fitting uncertainty to account for the absolute flux density
uncertainty as recommended for GLEAM sources at $-72^\circ \leq
\mathrm{Dec} \leq 18.5^\circ$ by~Hurley-Walker et al.~(submitted).\par

In case of non-detections, we used flux density upper limits in our study. Since
all sources are detected at 1.4\,GHz in the ATLAS survey at a confidence of at
least $9\sigma$ in DR3, all sources can be considered as unambiguous detections
at this frequency. Therefore, we used $3\sigma$ flux density upper limits in
case of non-detections at other wavelengths when using our own flux density
measurements. Since flux density upper limits of faint sources are dominated by
the local rms of the map and the calibration error hardly contributes, it is
valid for our study to neglect the fractional calibration error in case of
non-detections. For non-detections at 4.8\,GHz, 8.6\,GHz, and 34\,GHz, we used
the $3\sigma$ flux density upper limits quoted by \citet{Middelberg2011} and Emonts et al.~(in
prep.). The 20\,GHz catalogue from \citet{Franzen2014} contains sources with S/N
higher than 5. Therefore, we used $5\sigma$ flux density upper limits for
undetected sources at 20\,GHz. In case of non-detections at 200\,MHz in the
GLEAM survey, we set flux density upper limits as discussed in
Sects.~\ref{data_ELAIS_GLEAM} and \ref{data_CDFS_GLEAM}.\par

\subsection{Effects of different angular resolutions on our analysis}
\label{resolutioneffects}
In our analysis, we were using data covering a wide frequency range and taken
with different telescopes as described in Sects.~\ref{data_ELAIS} and
\ref{data_CDFS}. These observations therefore cover a wide range of resolution,
from a few arcsec to more than 100\,arcsec. We carefully checked that
our analysis is not affected by resolution effects.\par

Most sources from our sample are lacking complex structure and are point-like at
any frequency, so there are no significant resolution effects. However, flux
densities measured from lower-resolution maps can be increased because of
confusing, nearby radio sources. We checked all photometric detections for
potentially confusing radio sources---detected at higher resolution in the
610\,MHz and 1.4\,GHz observations---that might be located in the respective
beam covering the IFRS. If a measured flux density is or might be affected by
confusion, we did not use this data point in our analysis but considered it as a
flux density upper limit. We found potential issues in the 150\,MHz map, the
325\,MHz map, the 843\,MHz maps, the 844\,MHz map, and the 2.3\,GHz map and
discarded one, one, one, four, and three detections, respectively.\par

Particular caution had to be used with respect to the GLEAM counterparts because
of the large beam size. We found seven GLEAM counterparts of IFRS to be
potentially confused by other sources inside the GLEAM beam that are visible in
the higher-resolution data at 610\,MHz and 1.4\,GHz. GLEAM flux densities are
corrected for the local background, i.e.\ faint confusing sources of the order
of the local GLEAM rms do not contribute to the catalogued 200\,MHz flux
density. This was the case for three of these seven GLEAM counterparts. The
other four GLEAM counterparts, however, have strong closeby sources in the beam
and confusion is likely. Therefore, we used the GLEAM flux densities as upper
limits on the 200\,MHz flux densities of these four IFRS.\par

All flux density upper limits that were set because of confusion are specially
marked in Tables~\ref{tab:data_ELAIS} and \ref{tab:data_CDFS}. Our data might
also be affected by low surface brightness features that are measured at low
frequencies but are resolved out at higher frequencies. This would result in
decreasing high-frequency flux densities.\par

\subsection{Control sample}
\label{controlsample}
We built a control sample of the broader RL galaxy population---i.e.\
non-IFRS---to compare the results from our IFRS sample. For this, we randomly
selected 15~sources in ELAIS-S1 and 19~sources in CDFS, ensuring that they had
similar 1.4\,GHz flux densities than the IFRS from our sample described above.
Cross-matching with published source catalogues, measuring flux densities, and
dealing with flux density errors and confusion issues was carried out in the
same way as for the IFRS sample. However, since the observations at 4.8\,GHz,
8.6\,GHz, and 34\,GHz were targeted observations of IFRS, no data are available
for the sources in the control sample at these frequencies.\par

\subsection{PdBI observations and ancillary data of IFRS xFLS 478}
\label{data_xFLS478}
To complement the cm-wave observations described above, we observed one of the
brightest IFRS in the \citet{Zinn2011} catalogue, IFRS~xFLS\,478 (35.8\,mJy at
1.4\,GHz), with the PdBI. The source is located in the xFLS~\citep{Condon2003}
field at RA~17h11m48.526s and Dec +59d10m38.87s (J2000).
\citeauthor{Zinn2011} found an uncatalogued IR counterpart of around 20\,$\mu$Jy
at 3.6\,$\mu$m, resulting in a radio-to-IR flux density ratio
$S_{1.4\,\mathrm{GHz}}/S_{3.6\,\mu\mathrm{m}} = 1831$.\par

\subsubsection{PdBI observations}
\label{PdBI_observations}
The IFRS xFLS\,478 was observed in continuum with the PdBI at
105\,GHz~(2.9\,mm), covering a bandwidth of 3.6\,GHz. The observations were
carried out on 25-Aug-2013 and 13-Sep-2013 in 5Dq configuration and on
25-Sep-2013 and 02-Oct-2013 in 6Dq configuration. The field of view was
$51.2\times 51.2\,\mathrm{arcsec} ^2$ and the synthesised beam was $6.03\times
3.81\,\mathrm{arcsec} ^2$. The seeing varied between 0.95\arcsec and
2.44\arcsec. The data were correlated with the wide-band correlator WideX.\par

In all observations, MWC\,349 was observed as flux calibrator, while 1637+574
was used as phase and amplitude calibrator. Each of the four observing sessions
was divided into different scans. One scan consisted of 30 subscans of 45\,sec
each, corresponding to a total scan length of 22.5\,min. The phase and amplitude
calibrator were observed for 45\,sec after each scan on the target.\par

\subsubsection{PdBI data calibration, mapping, and flux measurement}
\label{PdBI_calibration}
Data calibration was carried out using the Grenoble Image and Line Data Analysis
Software\footnote{\url{http://www.iram.fr/IRAMFR/GILDAS}}~(GILDAS) packages. We
followed the different tasks in the \texttt{Standard Calibration} section of the
\texttt{CLIC} software included in GILDAS. Automatic flagging was applied and
phases were corrected for atmospheric effects. We measured the receiver bandpass
on 1803+784 (25-Sep-2013) or 3C\,454.3 (all other observing dates). In the
following, we departed from the standard calibration and calibrated phases and
amplitudes by averaging both polarisations, following a recommendation by the
PdBI staff. Phases and amplitudes were calibrated based on 1637+574, and
the flux density scale was then tied to MWC\,349.\par

Since the antenna configuration was changed immediately before the observations
on 25-Aug-2013, an incorrect baseline solution would have been used by the
standard calibration. Therefore, at the beginning of the data calibration, the most
suitable baseline solution---taken on 02-Sep-2013---was applied to these
data.\par

We performed a final flagging step on the calibrated data by flagging all
visibilities with phase losses $>40^\circ$ RMS or amplitude losses $>20$\% as
recommended for a detection experiment. 57\,116 visibilities remained after
this flagging process, corresponding to an effective on-source time on
IFRS xFLS\,478 of 11.9\,hrs with six antennas. A stricter phase loss
criterion in the flagging process did not improve our data.\par

Data analysis was done using the task \texttt{MAPPING} from the GILDAS software
package. We built the dirty image by applying natural weighting and using a
pixel size of 0.6\,arcsec and subsequently \texttt{CLEAN}ed the map.
\begin{figure}
	\centering
		\includegraphics[width=\hsize]{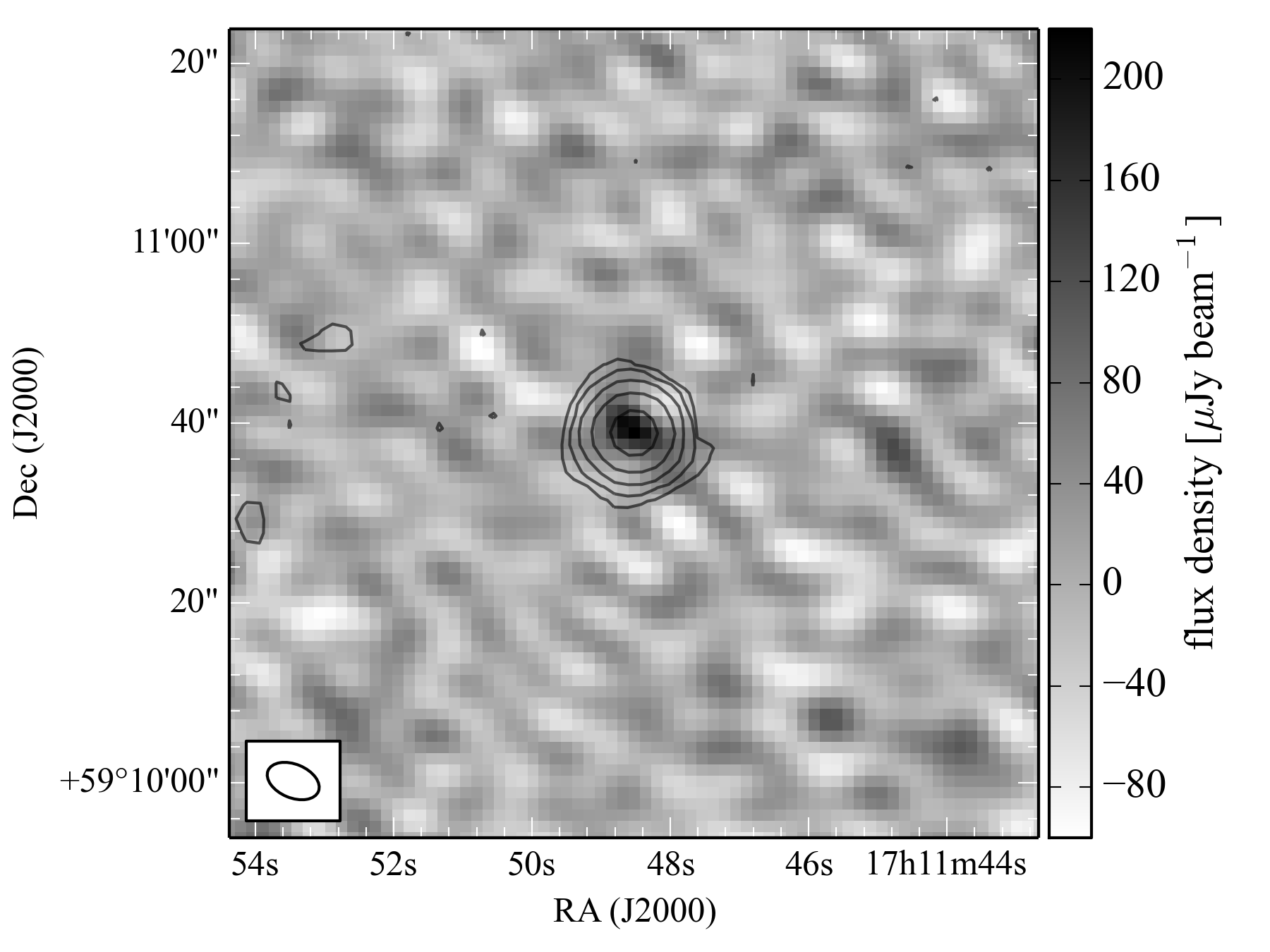}
		\caption{Plateau de Bure Interferometer map~(greyscale) of
		IFRS xFLS\,478 at 105\,GHz (2.9\,mm) overlaid with the VLA 1.4\,GHz radio
		contours from \citet{Condon2003}, starting at $3\sigma$ and increasing by
		factors of 4.}
	\label{fig:xFLS478map}
\end{figure}
The \texttt{CLEAN}ed map is shown in Fig.~\ref{fig:xFLS478map}.\par

Since the source appeared to be point-like in the \texttt{CLEAN}ed map, we
fitted the $uv$~data with the Fourier transform of a point source and found this fit to be
consistent. Based on the fit, we obtained a flux density of
220\,$\mu$Jy\,beam$^{-1}$ for xFLS\,478 at 105\,GHz. With a measured rms noise
of 36\,$\mu$Jy\,beam$^{-1}$, this corresponds to a $6.1\sigma$ detection. This is
the highest-frequency detection of an IFRS in the radio regime. The absolute
flux uncertainty is 10\%.\par

\subsubsection{Ancillary data of IFRS xFLS\,478}
\label{ancillarydata_xFLS478}
Counterparts of xFLS\,478 have been detected at 610\,MHz~(GMRT;
\citealp{Garn2007}), 325\,MHz~(Westerbork Northern Sky Survey;
\citealp{Rengelink1997}), and 151\,MHz~(6th Cambridge Survey;
\citealp{Hales1990}). In the near- and mid-IR regime, xFLS\,478 was observed
with \textit{Spitzer} and and detected at 4.5\,$\mu$m, but remained undetected
at 3.6\,$\mu$m, 5.8\,$\mu$m, and 8.0\,$\mu$m~\citep{Lacy2005}. Furthermore, the
source xFLS\,478 was observed by the \textit{Herschel} Multi-tiered
Extragalactic Survey~(HerMES; \citealp{Oliver2012}) and was detected at
250\,$\mu$m, 350\,$\mu$m, and 500\,$\mu$m. Source xFLS\,478 remained undetected
in the Sloan Digital Sky Survey data release~10 (SDSS DR10; \citealp{Ahn2014})
and also in the $R$~band survey~(50\% completeness at $24.5$\,Vega mag;
\citealp{Fadda2004}) with the Mosaic-1 camera on the Kitt Peak National
Observatory. Hence, the redshift of this source is unknown.\par

\section{Building and fitting the radio SEDs}
\label{building_fitting_radioSEDs}
Using the data presented in Sects.~\ref{data_ELAIS} and \ref{data_CDFS}, we
built radio SEDs for all 34~IFRS from our sample in CDFS and ELAIS-S1 based on
all photometric detections and flux density upper limits. The resulting radio
SEDs are shown in Fig.~\ref{fig:radioSEDs}.\par
\begin{figure*}
	\centering
		\includegraphics[width=17cm]{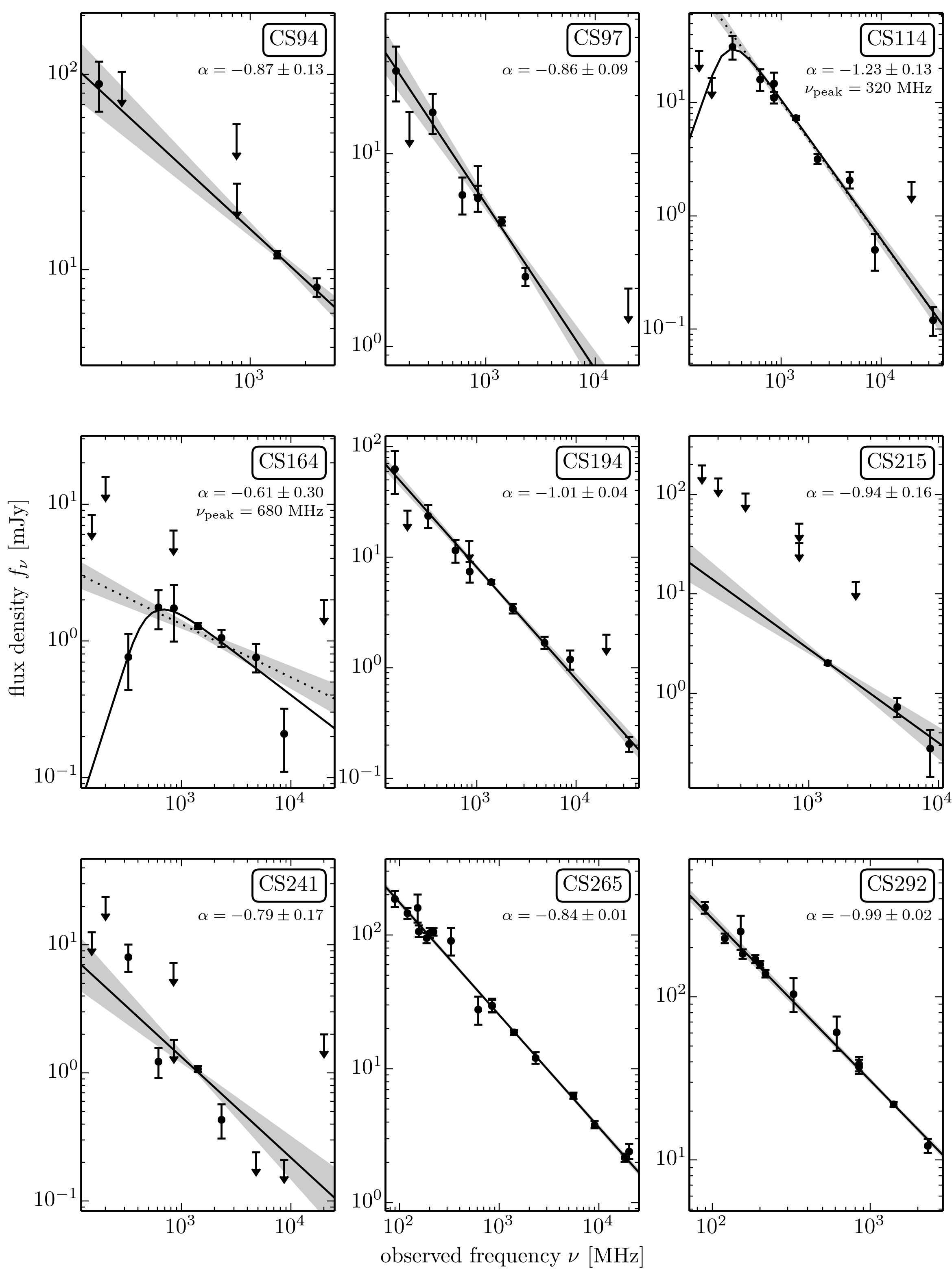}
		\caption{Radio SEDs of IFRS in CDFS and ELAIS-S1, using all available flux
		density data points and upper limits. The solid line shows the fit which was
		found to best describe the photometric detections as discussed in
		Sect.~\ref{building_fitting_radioSEDs}. Spectral index and---if applicable---turnover
		frequency of the best fit are quoted. We also show the first approach to
		describe the data---a single power law fitted to all photometric
		detections---by a dotted line if this fit was discarded later in the analysis.
		$1\sigma$ uncertainties of the single power law fits are represented by the
		shaded areas. Error bars show $1\sigma$ uncertainties. The frequency
		coverage varies from one IFRS to another and the flux density scales are
		different.}
		 \label{fig:radioSEDs}
\end{figure*}

\begin{figure*}
\ContinuedFloat
\centering
		\includegraphics[width=17cm]{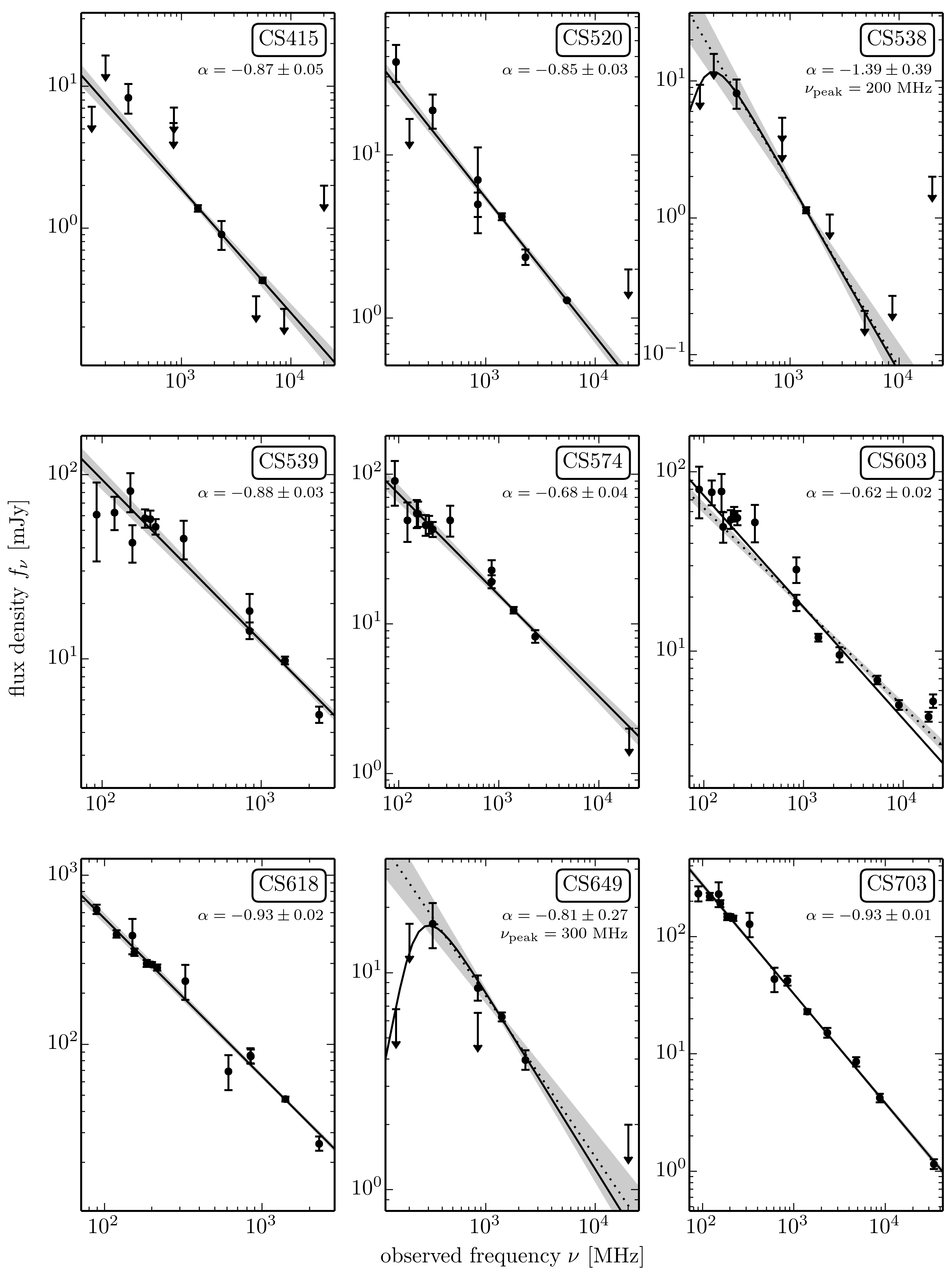}
	\caption{continued.}
\end{figure*}

\begin{figure*}
\ContinuedFloat
\centering
		\includegraphics[width=17cm]{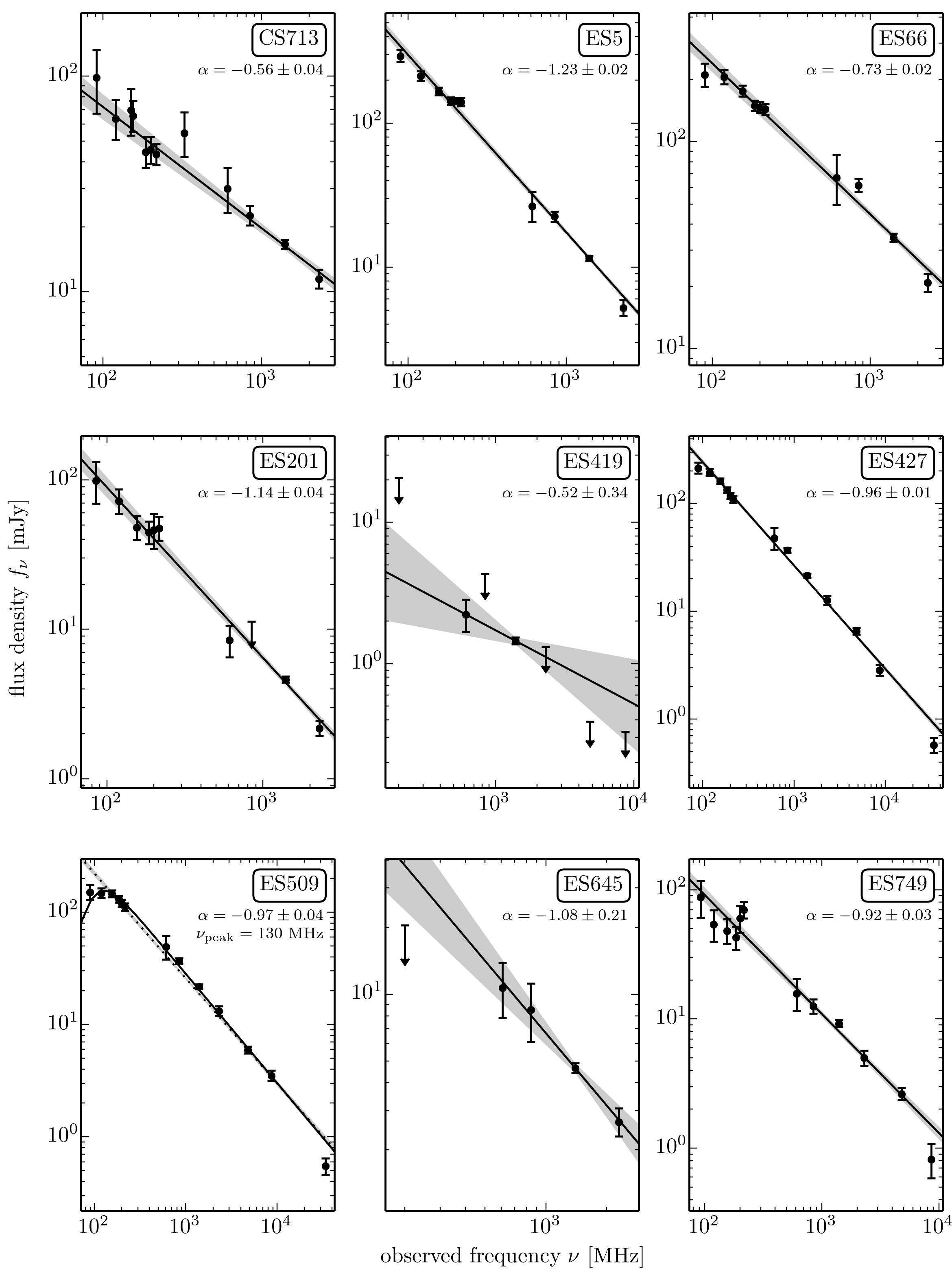}
	\caption{continued.}
\end{figure*}

\begin{figure*}
\ContinuedFloat
\centering
		\includegraphics[width=17cm]{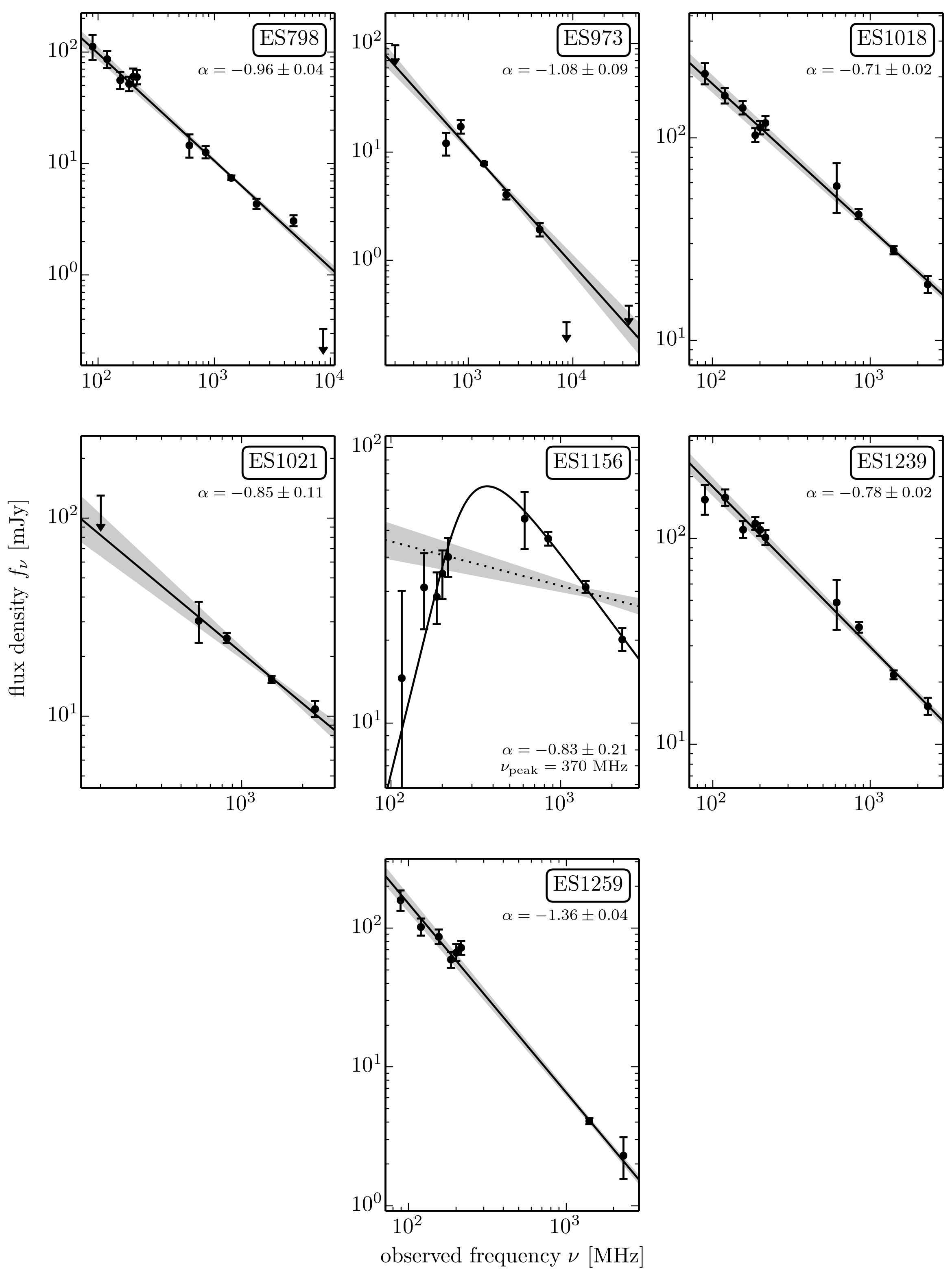}
	\caption{continued.}
\end{figure*}

\begin{enumerate}[(1)]

\item For each source, as the simplest approach, we fitted a single power law
based on a least-squared method to all available photometric detections of the
radio SED, weighting the data by their respective uncertainties. The
ATLAS DR3 in-band spectral indices were not used in the entire fitting
approach. The resulting fitted single power laws are shown in
Fig.~\ref{fig:radioSEDs}.\par

We considered the fitted single power law as an appropriate description
of the radio SED if (I)~the low-frequency tail (below 1.4\,GHz)  and (II)~the
high-frequency (above 1.4\,GHz) tail of the radio SED---considering detections
and upper limits---were consistent with the fit, and (III)~no turnover was seen
in the central part of the radio SED. More precisely, for (I) and (II), we required
that the cumulative low-frequency (high-frequency) deviation was below
$1\sigma$ or the fractional low-frequency (high-frequency) deviation per data
point was below $0.3\sigma$.\par

\item If the single power law in (1)~was rejected because of (I) or
(III), we fitted a radio SED model with a turnover to the photometric detections based on
a least-squared method and weighting the data points by their uncertainties. The
different models explaining this turnover can be divided by the location of the
related physical process: internal or external to the synchrotron-emitting
region~\citep{Kellermann1966}. If an external process is thought to cause the
turnover, the physical process is expected to be free-free absorption by ionised
gas outside the radio-emitting region. However, if the physical process is
internal, synchrotron self-absorption~(SSA) in the synchrotron-emitting region
itself is usually assumed to cause the turnover. Since we found our data to
trace only---if at all---the turnover in the radio SEDs but not the slope
towards low frequencies~(see Fig.~\ref{fig:radioSEDs}), we were not able to
study the physical processes causing the turnover. Therefore, the decision which
model to use for the fit was not relevant for our study and did not change our
results. We decided to use an SSA model~(e.g.\ \citealp{Tingay2003}) given by
\begin{equation}
S_\nu = S_0 \bigg(\frac{\nu}{\nu_0}\bigg)^{-\frac{\beta - 1}{2}}
\frac{1-e^{-\tau_\nu}}{\tau_\nu}~,~~~~~\tau_\nu =
\bigg(\frac{\nu}{\nu_0}\bigg)^{-\frac{\beta + 4}{2}}~,
\label{eq:SSA}
\end{equation}
where $S_0$ denotes the zero flux density, $\nu$ the frequency, $\nu_0$ the
frequency where the synchrotron optical depth is equal to 1, $\beta$ the power law index
of the relativistic electron energy distribution, and $\tau_\nu$ the
frequency-dependent optical depth. If the low-frequency end of the radio SED was
constrained by flux density upper limits and these limits were inconsistent with
the fitted single power law, we included these limits in the fitting to obtain a
lower limit on the peak frequency since lower flux densities at low frequencies
will push the peak towards higher frequencies. Sources fitted by the SSA model
are discussed in Sect.~\ref{spectra_turnover}. The fitted SSA model is shown for
these sources in Fig.~\ref{fig:radioSEDs}.\par

\item Sources that were found to be poorly described by a single power
law because of (II) can be divided in two subclasses, depending on the departure of
the fitted power law from the SED:
\begin{enumerate}
  \item If the highest frequency data points departed upwards from the
  fitted single power law, the source was considered as showing an upturn in
  the radio SED. A new single power law was then fitted to the data of this
  source, while ignoring the deviating high-frequency data points, and this
  fit is also shown in Fig.~\ref{fig:radioSEDs}. These sources are discussed
  in Sect.~\ref{upturn_in_radiospectrum}.
  \item If high-frequency flux densities or upper limits were found to
  depart downwards from the fitted single power law, the source was considered to
  steepen towards higher frequencies. This subclass is
  discussed in Sect.~\ref{steepening_spectra}.
\end{enumerate}
\end{enumerate}
\par
The classification of each IFRS and the spectral index at the
high-frequency side of the synchrotron bump---obtained from the best fit as described
above---are summarised in Table~\ref{tab:sample}.
\begin{table*}
	\caption{Characteristics and results of our sample of 34~IFRS in ELAIS-S1 and CDFS.}
	\label{tab:sample}
 \centering
 \begin{tabular}{c c c c c c c}
 	\hline \hline 
  IFRS    & IAU designation               & RA           & Dec            & $S_\mathrm{1.4\,GHz} / S_\mathrm{3.6\,\mu m}$ & $\alpha$           & Classification    \\ 
  ID      &                               & J2000        & J2000          &                                             &                    &                   \\ \hline
  CS94    & ATCDFS\,J032740.72--285413.4  & 03:27:40.727 & --28:54:13.48  & $801  $                                     & $ -0.87 \pm 0.13 $ & pl, poss.\ peak   \\      
  CS97    & ATCDFS\,J032741.70--274236.6  & 03:27:41.700 & --27:42:36.61  & $614  $                                     & $ -0.86 \pm 0.09 $ & pl                \\ 
  CS114   & ATCDFS\,J032759.89--275554.7  & 03:27:59.894 & --27:55:54.73  & $>2400$                                     & $ -1.23 \pm 0.13 $ & peak              \\
  CS164   & ATCDFS\,J032900.20--273745.7  & 03:29:00.200 & --27:37:45.70  & $640  $                                     & $ -0.61 \pm 0.30 $ & peak, st           \\
  CS194   & ATCDFS\,J032928.59--283618.8  & 03:29:28.594 & --28:36:18.81  & $>2033$                                     & $ -1.01 \pm 0.04 $ & pl                \\
  CS215   & ATCDFS\,J032950.01--273152.6  & 03:29:50.010 & --27:31:52.60  & $>733 $                                     & $ -0.94 \pm 0.16 $ & pl, poss.\ peak   \\     
  CS241   & ATCDFS\,J033010.21--282653.0  & 03:30:10.210 & --28:26:53.00  & $>908 $                                     & $ -0.79 \pm 0.17 $ & pl, poss.\ peak   \\      
  CS265   & ATCDFS\,J033034.66--282706.5  & 03:30:34.661 & --28:27:06.51  & $634  $                                     & $ -0.84 \pm 0.01 $ & pl                \\   
  CS292   & ATCDFS\,J033056.94--285637.2  & 03:30:56.949 & --28:56:37.29  & $1842 $                                     & $ -0.99 \pm 0.02 $ & pl                \\  
  CS415   & ATCDFS\,J033213.07--274351.0  & 03:32:13.070 & --27:43:51.00  & $>1186$                                     & $ -0.87 \pm 0.05 $ & pl, poss.\ peak   \\  
  CS520   & ATCDFS\,J033316.75--280016.0  & 03:33:16.754 & --28:00:16.02  & $500  $                                     & $ -0.85 \pm 0.03 $ & pl                \\    
  CS538   & ATCDFS\,J033330.20--283511.1  & 03:33:30.200 & --28:35:11.10  & $>648 $                                     & $ -1.39 \pm 0.39 $ & peak              \\     
  CS539   & ATCDFS\,J033330.54--285428.2  & 03:33:30.542 & --28:54:28.22  & $640  $                                     & $ -0.88 \pm 0.03 $ & pl                \\        
  CS574   & ATCDFS\,J033353.27--280507.3  & 03:33:53.279 & --28:05:07.31  & $1091 $                                     & $ -0.68 \pm 0.04 $ & pl, poss.\ peak   \\        
  CS603   & ATCDFS\,J033413.75--283547.4  & 03:34:13.759 & --28:35:47.47  & $709  $                                     & $ -0.62 \pm 0.02 $ & upturn            \\       
  CS618   & ATCDFS\,J033429.75--271744.9  & 03:34:29.754 & --27:17:44.95  & $1660 $                                     & $ -0.93 \pm 0.02 $ & pl                \\          
  CS649   & ATCDFS\,J033452.84--275813.0  & 03:34:52.846 & --27:58:13.05  & $1838 $                                     & $ -0.81 \pm 0.27 $ & peak              \\     
  CS703   & ATCDFS\,J033531.02--272702.2  & 03:35:31.025 & --27:27:02.20  & $>8700$                                     & $ -0.93 \pm 0.01 $ & pl                \\        
  CS713   & ATCDFS\,J033537.52--275057.8  & 03:35:37.525 & --27:50:57.88  & $643  $                                     & $ -0.56 \pm 0.04 $ & pl                \\      
  ES5     & ATELAIS\,J003709.36--444348.1 & 00:37:09.365 & --44:43:48.11  & $1082 $                                     & $ -1.23 \pm 0.02 $ & pl                \\  
  ES66    & ATELAIS\,J003942.45--442713.7 & 00:39:42.452 & --44:27:13.77  & $1865 $                                     & $ -0.73 \pm 0.02 $ & pl                \\     
  ES201   & ATELAIS\,J003130.06--441510.6 & 00:31:30.068 & --44:15:10.69  & $>1683$                                     & $ -1.14 \pm 0.04 $ & pl                \\   
  ES419   & ATELAIS\,J003322.76--435915.3 & 00:33:22.766 & --43:59:15.37  & $557  $                                     & $ -0.52 \pm 0.34 $ & pl, poss.\ peak   \\  
  ES427   & ATELAIS\,J003411.59--435817.0 & 00:34:11.592 & --43:58:17.04  & $>7120$                                     & $ -0.96 \pm 0.01 $ & st                \\  
  ES509   & ATELAIS\,J003138.63--435220.8 & 00:31:38.633 & --43:52:20.80  & $>7400$                                     & $ -0.97 \pm 0.04 $ & peak, st          \\     
  ES645   & ATELAIS\,J003934.76--434222.5 & 00:39:34.763 & --43:42:22.58  & $780  $                                     & $ -1.08 \pm 0.21 $ & pl, poss.\ peak   \\    
  ES749   & ATELAIS\,J002905.22--433403.9 & 00:29:05.229 & --43:34:03.94  & $>2337$                                     & $ -0.92 \pm 0.03 $ & st                \\       
  ES798   & ATELAIS\,J003907.93--433205.8 & 00:39:07.934 & --43:32:05.83  & $>2597$                                     & $ -0.96 \pm 0.04 $ & st                \\      
  ES973   & ATELAIS\,J003844.13--431920.4 & 00:38:44.139 & --43:19:20.43  & $>3046$                                     & $ -1.08 \pm 0.09 $ & st, poss.\ peak   \\ 
  ES1018  & ATELAIS\,J002946.52--431554.5 & 00:29:46.525 & --43:15:54.52  & $1012 $                                     & $ -0.71 \pm 0.02 $ & pl                \\
  ES1021  & ATELAIS\,J003255.53--431627.1 & 00:32:55.534 & --43:16:27.15  & $575  $                                     & $ -0.85 \pm 0.11 $ & pl, poss.\ peak   \\
  ES1156  & ATELAIS\,J003645.85--430547.3 & 00:36:45.856 & --43:05:47.39  & $2888 $                                     & $ -0.83 \pm 0.21 $ & peak              \\
  ES1239  & ATELAIS\,J003547.96--425655.4 & 00:35:47.969 & --42:56:55.40  & $1220 $                                     & $ -0.78 \pm 0.02 $ & pl                \\
  ES1259  & ATELAIS\,J003827.17--425133.7 & 00:38:27.170 & --42:51:33.70  & $>2063$                                     & $ -1.36 \pm 0.04 $ & pl                \\ \hline
	\end{tabular}
	\tablefoot{The IAU designations and positions are taken from \citet{Norris2006} and \citet{Middelberg2008ELAIS-S1}, radio-to-IR flux density ratios from \citet{Zinn2011} and \citet{Maini2013submitted}. Spectral indices~$\alpha$ and classification of the radio SEDs are results of our work as described in Sect.~\ref{building_fitting_radioSEDs}. If a source was found to show a turnover, it is classified by ``peak'' and the quoted spectral index was obtained from fitting the SSA model to the data. Sources with data that are well described by a single power law are labelled as ``pl''. The additional classification ``poss.\ peak'' indicates sources for which a turnover cannot be ruled out down to 200\,MHz, mainly because of their faintness. Sources steepening towards higher frequencies are labelled as ``st''. Sources with increasing flux densities at the highest frequencies are indicated by ``upturn''. Spectral indices were measured over different frequency ranges as discussed in Sect.~\ref{building_fitting_radioSEDs}.}
\end{table*}
Also listed is the IAU designation, the position, and the radio-to-IR
flux density ratio from \citet{Zinn2011} or \citet{Maini2013submitted}. We
do not quote reduced chi-squared numbers for the fits since
upper limits were used as constraints in some cases as discussed above. A
statistical comparison between the fits based on these numbers would be
incorrect.\par

The SEDs of the sources in our control sample were built and fitted in the same
way. We found the SEDs to be self-consistent, i.e.\ without spectral features
that might arise from flux density measurements at different angular
resolutions.
Since the IFRS and control samples would suffer from the same effects, we are
confident that our analysis is not significantly affected by changing
resolution. In particular, we found that our approach to classify radio SEDs as
described above works for the IFRS sample and for the control sample. In the
subsequent analysis, we quote numbers for the control sample in square
brackets.\par

The class of IFRS has not been studied with respect to radio variability.
Therefore, variability effects on the radio SEDs presented here cannot be ruled
out. In general, long-term variability (of the order of a year) of radio sources
is low at 1.4\,GHz and lower frequencies~(e.g.\
\citealp{Ofek2011,Thyagarajan2011,Mooley2013}). However, this is not necessarily
the case at higher frequencies $\gtrsim 5$\,GHz where a significant fraction---a
few tens per cent---of sources show variability of the order of 10\% or
more~(e.g.\ \citealp{Bolton2006,Sadler2006,Franzen2009,Chen2013}). In
particular, flat- or inverted-spectrum radio sources are variable because of
their dominating, beamed core emission~(e.g.\ \citealp{Franzen2014}). These
classes of object usually dominate samples selected at $\sim 20$\,GHz.
So it is very unlikely that the 1.4\,GHz flux densities of our sample
are significantly affected by variability, but we have no information about
variability at higher frequencies.\par

\section{Discussion: Radio SEDs of IFRS}
\label{discussion_radioSEDs}

\subsection{Sources following a single power law}
\label{powerlaw}

Out of our sample of 34~IFRS [34~sources in the control sample], the SEDs of
23~IFRS [29~sources from the control sample] were well described by a single
power law fitted to all available photometric data as described in
Sect.~\ref{building_fitting_radioSEDs}. These sources do not show any evidence
for a deviation from this fit, neither at low nor high frequencies. However, we
note that nine~[eight] of these sources are comparatively faint or are affected
by confusion in some of the observations, reducing the number of photometric
detections and, consequently, the number of data points constraining their radio
SEDs. Therefore, we were able to exclude a deviation from the fitted single
power law---by increasing or decreasing flux density at low or high
frequencies---for only 14 of the 23~IFRS or $61^{+9}_{-11}$\%~[21 of the
29~sources in the control sample, or $72^{+7}_{-10}$\%] based on the available
data.\par

\citet{Klamer2006} studied the radio SEDs of a sample of 37~HzRGs,
selected at observed frequencies between 843\,MHz and 1.4\,GHz. The majority of their
sources (89\%) were found to be well described by a single power law in the
studied frequency range between 843\,MHz and 18\,GHz. Our frequency coverage
extends significantly to lower frequencies compared to theirs. If considering
only the radio SEDs above 800\,MHz, we found $82^{+5}_{-8}$\% of our
IFRS to be well described by a single power law, similar to the HzRG sample from
\citeauthor{Klamer2006} \citet{Emonts2011a} and \citet{Emonts2011b} found three
HzRGs to follow single power laws up to frequencies of 36\,GHz. Out of
the five IFRS with 34\,GHz detections presented in our work, we find three
to be consistent with a single power law up to 34\,GHz, whereas the SEDs of the
remaining two IFRS slightly steepen towards that frequency.\par

\subsection{Radio spectral index}
\label{spectral_index}
Based on the best fit found for each IFRS as described in
Sect.~\ref{building_fitting_radioSEDs}, we found spectral indices
between $-0.52$ and $-1.39$~[between $-0.01$ and $-1.6$] on the high-frequency
side of the synchrotron bump for the 34~IFRS in our sample. The median index is
$-0.88\pm 0.04$~[$-0.74 \pm 0.06$] and the mean index is $-0.91 \pm
0.20$~[$-0.69 \pm 0.29$]. We emphasise that more high-frequency data is
available for the IFRS sample than for the control sample. Therefore, numbers of
the two samples cannot be compared.\par

Our median spectral index for IFRS of $-0.88$ is flatter than the median index
of $-1.4$ for IFRS found by \citet{Middelberg2011}. However, we measured the
spectral index over a wider frequency range---particularly towards lower
frequencies---, whereas the median index from \citeauthor{Middelberg2011} has
been measured between 1.4\,GHz and 2.4\,GHz. \citeauthor{Middelberg2011} also
find a spectral steepening towards higher frequencies which is discussed in
detail in Sect.~\ref{steepening_spectra}. They present a median spectral index
for HzRGs of $\alpha = -1.02$ between 1.4\,GHz and 2.4\,GHz which is close to
the number found in our study for IFRS. Our median spectral index is steeper
than the median spectral index of the entire radio source population ($\alpha =
-0.74$) and the AGN population ($\alpha = -0.63$) in the ATLAS fields as
presented by \citet{Zinn2012} between 1.4\,GHz and 2.3\,GHz. The median spectral
index of the broader radio source population presented by \citeauthor{Zinn2012}
is consistent with the median spectral index of $-0.74$ found for our control
sample.\par

\begin{figure}
	\centering
		\includegraphics[width=\hsize]{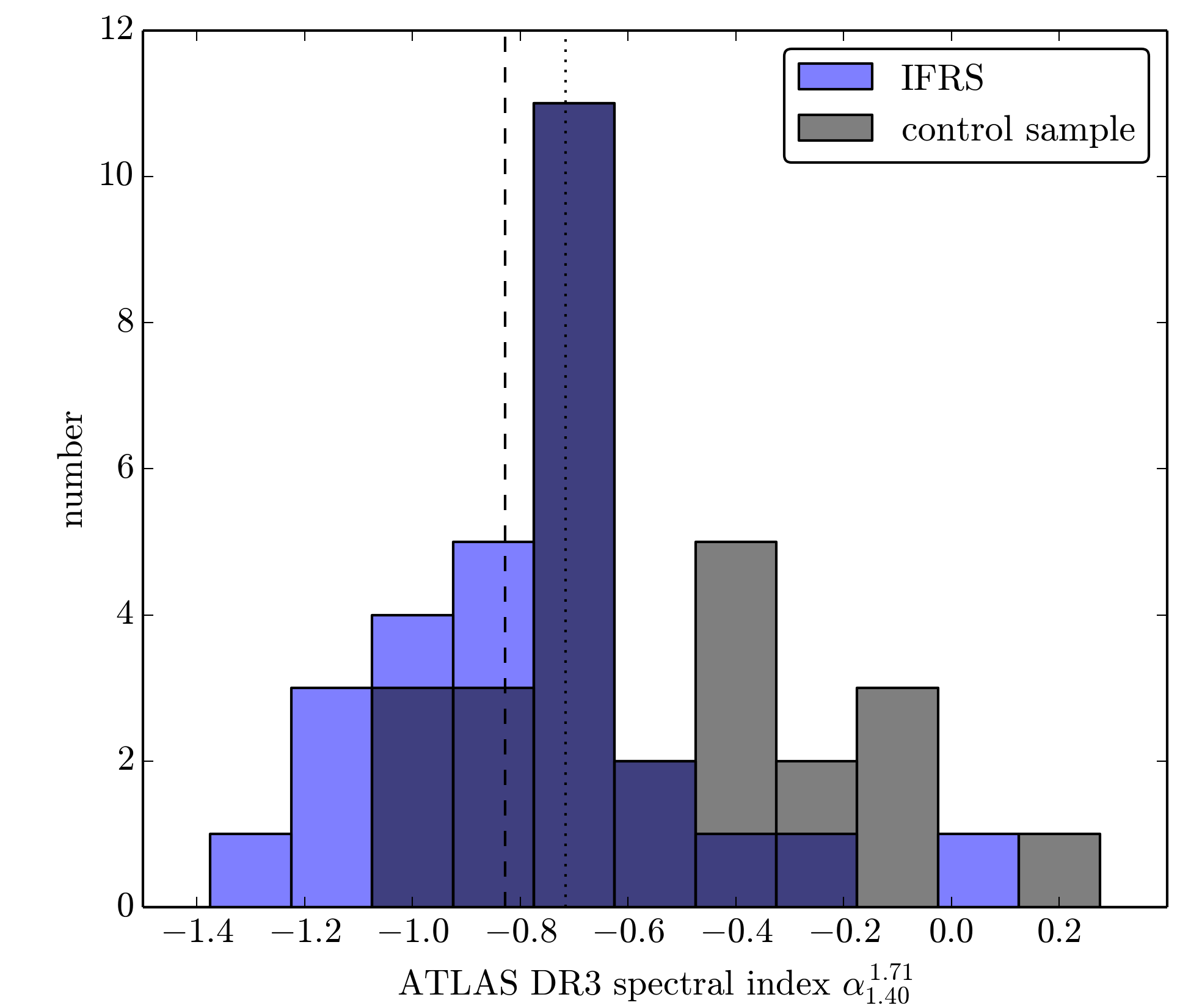}
		\caption{Histogram of the ATLAS DR3 spectral indices~$\alpha^{1.71}_{1.40}$,
		described in Sects.~\ref{data_ELAIS_1p4GHz} and \ref{data_CDFS_1p4GHz} and
		quoted in Tables~\ref{tab:data_ELAIS} and \ref{tab:data_CDFS}.
		The IFRS sample is represented by blue bars and the control sample by grey
		bars. The vertical lines show the median spectral indices of the IFRS sample
		($-0.83$; black dashed line) and the control sample ($-0.72$; black dotted
		line).}
	\label{fig:histogram_specindex}
\end{figure}

We also studied the 1.4\,GHz spectral indices from the ATLAS DR3 data.
In contrast to the spectral indices obtained from a fit to the radio SED described
above, these 1.4\,GHz spectral indices are based on the same data for the IFRS
sample and the control sample. Therefore, both samples can be properly compared
based on the 1.4\,GHz spectral indices. The histogram of these spectral indices
is shown in Fig.~\ref{fig:histogram_specindex}.
We found that the IFRS sample has a steeper median radio SED than the control
sample and that the spectral index distribution of IFRS is shifted towards steeper SEDs
compared to the control sample, describing the broader, flux density-matched
radio source population. The intrinsic difference between these two populations
is also shown by a two-sample Anderson-Darling~(A-D) test~\citep{Scholz1987}.
The A-D~test measures the sum of the squared deviations of the samples and is more
sensitive than a Kolmogorov-Smirnov~(K-S), in particular at the tails of the
distribution~\citep{Babu2006}. We rejected the null hypothesis that the spectral
indices in the IFRS sample and in the control sample have the same parent distribution
(probability $p < 0.0015$).\par

\subsection{Ultra-steep, steep, flat, and inverted radio SEDs}
\label{ultrasteep_spectra}
The IFRS in our sample show generally steep radio SEDs. However, there is no
generally accepted definition for steep and ultra-steep spectrum~(USS) sources
and selection criteria differ between studies with respect to frequencies and
critical spectral index. Steep radio SEDs might be defined based on a spectral
index~$\alpha < -0.8$. Following this criterion, 25 ($74^{+6}_{-9}$\%)
[7; $21^{+9}_{-5}$\%] out of 34~IFRS can be classified as steep-spectrum
sources.\par

\citet{Afonso2011} point out that a significant number of sources with
measured spectral indices steeper than a critical index are likely to be intrinsically
flatter due to the long tail of the spectral index distribution. They argue that
$\alpha < -1.0$ is a reasonable definition for USS sources and used a
conservative cut $\alpha < -1.3$ between 610\,MHz and 1.4\,GHz for their sample.
We found two IFRS ($6^{+7}_{-2}$\%; CS538, ES1259) [1;
$3^{+6}_{-1}$\%] in our sample with a spectral index steeper than $-1.3$. They
are most likely to be USS sources. Further, six~($18^{+8}_{-5}$\%) [3;
$9^{+7}_{-3}$\%] IFRS have a spectral index in the range $-1.3 \leq \alpha \leq
-1.0$. These sources are also good candidates for USS sources. In particular,
we found statistically significantly more steep-spectrum sources in the IFRS sample
than in the control sample. Based on a Fisher's exact test~(e.g.\
\citealp{Wall2012}), we found a probability~$p < 0.001$ that the
subsets of sources with $\alpha < -0.8$ in the IFRS and the control samples were
obtained from the same parent spectral index distribution, consistent with the
different ATLAS spectral index distributions presented in
Sect.~\ref{spectral_index}.\par

USS sources in IFRS samples were already found by \citet{GarnAlexander2008} who
classify three~($21^{+14}_{-7}$\%) IFRS in their sample as USS sources based
on a spectral index $\alpha < -1$ between 610\,MHz and 1.4\,GHz.
\citet{Collier2014} find 155~($16^{+1}_{-1}$\%) USS sources---defined by
$\alpha \lesssim -1.0$---in their all-sky sample of IFRS.\par

Steep-spectrum radio sources tend to be at higher redshifts~(e.g.\
\citealp{Tielens1979,McCarthy1991,Roettgering1994,Chambers1996,Klamer2006}),
although exceptions in both directions are known~(see references in
\citealp{Afonso2011}). The suggested high redshifts of IFRS---the highest known
redshift is $z=2.99$~\citep{Collier2014}---have been confirmed based on optical
spectroscopy~\citep{Collier2014,Herzog2014}. However, it has been argued that
the IR-faintest IFRS might be at even higher
redshifts~\citep{Norris2011,Collier2014,Herzog2014}. Our finding that the
fraction of steep-spectrum sources is higher in the IFRS sample than in the
control sample can be interpreted that IFRS might be at higher redshifts than
ordinary RL AGN.\par

Two IFRS from our sample have spectroscopic redshifts: CS265~($z=1.84$) and
CS713~($z=2.13$). They are among the IR and optically brightest IFRS in the
ATLAS fields and are therefore expected to be at the lower tail of the redshift
distribution of IFRS. Following the connection between steepness of the radio
SED and redshift, the radio spectral indices for these sources presented in this
work of $-0.84$ and $-0.56$---lower than the median spectral index---suggest
that these sources have lower redshifts than the median IFRS in our sample,
consistent with the argument based on the IR flux densities.\par

We found two~($6^{+7}_{-2}$\%) IFRS in our sample [10;
$29^{+9}_{-7}$\%] with a flat ($-0.6 \leq \alpha \leq 0$) and none ($0^{+5}$\%)
[0; $0^{+5}$\%] with an inverted ($\alpha > 0$) radio SED. Based on these numbers, we are confident
that the radio SEDs presented in this work are not significantly affected by
radio variability as discussed in Sect.~\ref{building_fitting_radioSEDs}.\par

\subsection{Radio SEDs steepening towards higher frequencies}
\label{steepening_spectra}
We found five IFRS ($15^{+8}_{-4}$\%; CS164, ES427, ES509, ES798, ES973)
[one; $3^{+6}_{-1}$\%] in our sample that show a steepening radio SED towards
higher frequencies, suggesting that a single power law does not properly describe the
data. This spectral behaviour was already found for two of these IFRS by
\citet{Middelberg2011} and can be explained by a recently inactive AGN. In a
magnetic field, higher-energy electrons lose their energy faster than low-energy
electrons. If a region of synchrotron emission is not fed by the continuous
injection of new particles, the highest-energy particles are cooled quicker by
energy losses than low-energy particles, resulting in a lack of radiated
high-energy photons and a steepening in the SED towards higher
frequencies~(e.g.\ \citealp{Kardashev1962}).\par

\citet{Middelberg2011} matched the $uv$~coverage of their observations at
4.8\,GHz and 8.6\,GHz to eliminate the possibility that the observed spectral
steepening between 4.8\,GHz and 8.6\,GHz might be caused by resolution effects.
Therefore, this can be ruled out for ES798 and ES973---both detected at 4.8\,GHz
but undetected at 8.6\,GHz---and CS164 which was detected at both
frequencies.\par

Based on their resolution-matched spectral indices between 1.4\,GHz and 2.4\,GHz
on the one hand and between 4.8\,GHz and 8.6\,GHz on the other hand,
\citet{Middelberg2011} find that the radio SEDs of IFRS generally steepen
towards higher frequencies. Some IFRS in our sample are also steepening towards
higher frequencies. However, our data at high frequencies are generally not
sensitive enough to detect or constrain the radio SED of our IFRS. Only four
IFRS were detected in the higher-frequency surveys by \citet{Huynh2012} and
\citet{Franzen2014}; these are the only IFRS from our sample detected at a
frequency above 2.3\,GHz that were not covered in the observations by
\citeauthor{Middelberg2011} For these four sources, we did not find evidence for
a steepening. In contrast, one of those four IFRS even shows an upturn as
discussed in the following Sect.~\ref{upturn_in_radiospectrum}.\par

We found a steepening radio SED towards higher frequencies for
only one source in our control sample. However, few high-frequency
data are available in the control sample since the observations at 4.8\,GHz,
8.6\,GHz, and 34\,GHz were targeted observations of IFRS and no data are
available at these frequencies for the sources in the control sample. Therefore,
we cannot exclude the possibility that a steepening occurs for some of the
control sources but is not seen in our data because of poor high-frequency
coverage and sensitivity. \citet{Klamer2006} did not find any HzRG in their
sample of 37~sources that steepens at higher frequencies. If the fraction of
steepening sources is higher for IFRS than for HzRGs, this might suggest an
intrinsic difference. In that case, IFRS might be recently inactive and
restarted RL~AGN, whereas HzRGs do not show any evidence for a changing
activity of their active nucleus.\par

\subsection{IFRS with an upturn in their radio SED}
\label{upturn_in_radiospectrum}
The radio SED of IFRS CS603 follows a single power law in the frequency range
between 800\,MHz and 10\,GHz. At higher frequencies, however, the SED departs
from this power law, showing an increasing flux density with increasing
frequency. This is indicated by the 18\,GHz detection and clearly visible from
the 20\,GHz detection. There are two potential explanations for this
behaviour.\par

A flattening or upturning SED at high frequencies can be explained by a flat or
inverted SED of an AGN core that is dominating over the steep synchrotron SED of
the lobes at these frequencies. Alternatively, the upturn might be caused by
dust. It is known that thermal free-free and dust emission start to dominate
over the non-thermal synchrotron emission at rest-frame frequencies above $\sim
100$\,GHz in starburst galaxies~(e.g.\ \citealp{Murphy2009}; Fig.~2), though
thermal dust emission significantly depends on the size and composition of the
dust grains. Considering that IFRS are known to be AGN and that no evidence for
heavy dust obscuration in IFRS has been found~\citep{Collier2014}, the flat or
inverted radio SED of an AGN core seems to be the most plausible explanation.
This interpretation is consistent with results from
\citet{Hogan2015a,Hogan2015b}, who find that a fainter, flatter spectral core
component is often present in the radio SEDs of brightest cluster
galaxies~(BCGs). However, higher-frequency observations of IFRS CS603 are needed
to add evidence to this hypothesis.\par

Since we found an upturning SED at high frequencies also for one source
in the control sample, this spectral behaviour does not seem to be a characteristic
feature of IFRS, but to occur in the broader RL~AGN population, too. The results
from \citet{Klamer2006}, finding 11\% of their HzRGs flattening at higher
frequencies, is also consistent. The putative causes for this effect discussed
with respect to the IFRS are also valid for HzRGs and ordinary AGN without
IR-faintness.\par

\subsection{Radio SEDs showing a turnover}
\label{spectra_turnover}

Covering the frequency regime between 200\,MHz and 34\,GHz in ELAIS-S1 and
between 150\,MHz and 34\,GHz in CDFS, our data enabled us to detect the turnover
in the radio SEDs of IFRS in a wide frequency range. In particular, GPS sources
with a turnover frequency above 500\,MHz and CSS sources, peaking at frequencies
below 500\,MHz, should be detectable based on our rich data set. It has been
argued by \citet{Middelberg2011} and \citet{Herzog2015a} that an overlap between
the population of GPS and CSS on the one hand and IFRS on the other hand might
exist. \citet{Collier2014} find that at least a few IFRS are GPS or CSS
sources.\par

The radio SEDs shown in Fig.~\ref{fig:radioSEDs} revealed that CS164,
ES509, and ES1156 have a turnover in the frequency range of a few
hundred MHz [three sources in the control sample]. Based on fitting an SSA model
to the data, we found peak frequencies in the observed frame between
130\,MHz and 680\,MHz [200\,MHz -- 1.1\,GHz]. In addition to these
three IFRS, the radio SEDs of CS114, CS538, and CS649 [no source] also
suggest a turnover in the frequency regime covered by our data. However, the putative peak in the
radio SEDs of these three sources is indicated only by flux density upper
limits. Based on fitting the SSA model to the flux density upper limits, we
obtained lower limits of the peak frequencies between 200\,MHz and
320\,MHz for these three sources.\par

Summarising, we found six~($18^{+8}_{-5}$\%) IFRS [3; $9^{+7}_{-3}$\%]
in our sample of 34~sources that show a clear turnover in their radio SED based
on photometric detections or flux density upper limits. Out of these peaking
sources, one IFRS [one] was found to peak at an observed frequency above
500\,MHz, fulfilling the selection criterion of GPS sources (note that GPS
sources are usually defined based on their observed peak frequency;
\citealp{ODea1998}). Based on these numbers, we suggest that $3^{+6}_{-1}$\% of
IFRS [$3^{+6}_{-1}$\%] are GPS sources. Considering that we cannot rule out a
turnover in the frequency range above 200\,MHz for nine [eight] other sources
since their low-frequency regime is only constrained by upper limits, we
conclude that between $18^{+8}_{-5}$\% and $44^{+9}_{-8}$\% of IFRS [between
$9^{+7}_{-3}$\% and $32^{+9}_{-7}$\%] show a turnover at a frequency above $\sim
150$\,MHz. However, since CSS sources can also have their turnover at
frequencies below 150\,MHz---i.e.\ even IFRS following a single power law down
to 150\,MHz might be CSS sources---, we are not able to set an upper limit on
the fractional overlap between IFRS and CSS sources. However, we suggest that
this overlap is $\geq 15^{+8}_{-4}$\% [$\geq 6^{+7}_{-3}$\%]. The class of CSS
sources~(e.g.\ \citealp{ODea1998}) is defined by steep radio SEDs~($\alpha
\lesssim -0.5$) and compact morphology (a few or a few tens of kpc). Since IFRS
are known to be compact with linear sizes of not more than a few tens of kpc
(e.g.\ \citealp{GarnAlexander2008,Middelberg2011}) and to have steep radio SEDs
(\citealp{Middelberg2011}; and discussion in Sects.~\ref{spectral_index} and
\ref{ultrasteep_spectra}), IFRS are prototypical for the class of CSS sources.
In particular, we found five additional sources (CS415, CS539, ES66,
ES427, and ES645) that slightly departed from the fitted single power law or
flattened at low frequencies and might be CSS sources not represented in our
statistics. Therefore, we suggest that the fraction of CSS sources is
putatively significantly higher than the observed fraction of
$15^{+8}_{-4}$\%.\par

In our control sample, we found $3^{+6}_{-1}$\% and $\geq 6^{+7}_{-3}$\% of the
sources to be GPS sources and CSS sources, respectively. Considering
that the latter number is a lower limit, these numbers are consistent with
those found by \citet{ODea1998} in the broader population of RL AGN ($\sim 10$\%
and $\sim 30$\%, respectively). Comparing these numbers to those found for our
IFRS sample, we did not find evidence for a higher fraction of GPS and CSS
compared to samples of ordinary RL~AGN. However, the potentially high redshifts
of our IFRS sample might prevent us from tracing the expected peak in the radio
SED covered by our data. Although suffering from small number statistics, the
lower turnover frequencies found in the IFRS sample compared to the control
sample is consistent with putatively higher redshifts of IFRS.\par

When comparing samples selected at different frequencies, selection
biases have to be taken into account. Generally, the selection frequencies
affect the number of detected sources in the respective samples. When comparing
the fractions of GPS sources, an additional selection bias is present. Since the
low-frequency slope in the radio SED of a GPS source is usually steeper than the
high-frequency slope, a sample selected at 1.4\,GHz is more likely to find
sources peaking at lower frequencies than samples selected at higher frequencies
like the Tenth Cambridge~(10C) survey of radio
sources~\citep{AMI_Davies2011,AMI_Franzen2011,Whittam2015} at 15.7\,GHz or the
AT20G~\citep{Murphy2010,Franzen2014} survey at 20\,GHz. This provides a
potential explanation both for the lower GPS fraction in our sample compared to
the sample presented by \citet{ODea1998}---mainly selected at 5\,GHz---and for
the finding that the few GPS sources in our sample are peaking at relatively low
frequencies.\par

It has already been argued that a significant fraction of IFRS might be young
AGN in their earliest evolutionary stages~\citep{Collier2014,Herzog2015a}.
Although our results based on the turnover do not provide evidence for a higher
fraction of GPS and CSS sources in the IFRS population, they do not exclude this
possibility either. Instead, a high fraction of CSS sources is likely because of
the steep radio SEDs and the compact morphology of IFRS. If IFRS are indeed
younger---i.e.\ with turnovers at high rest-frame frequencies---and at higher
redshifts than the broader AGN population, these two effects would work against
each other. A younger radio galaxy is expected to peak at a higher rest-frame
frequency, but a high redshift shifts this peak to a lower frequency, resulting
in similar observed fractions of GPS and CSS sources in the IFRS population and
in the broader RL AGN population. Since we are lacking redshifts for the vast
majority of IFRS in our sample, we are unable to distinguish between these two effects
contributing to the observed peak frequency: evolution of the AGN, and
cosmology.\par

\subsection{Connection between turnover frequency and linear size}
\label{connection_turnover_linearsize}

Based on samples of GPS and CSS sources, \citet{ODea1997} presents an
anti-correlation between intrinsic turnover frequency and linear size~(their
Fig.~3). In the evolutionary scenario for AGN described in
Sect.~\ref{spectra_turnover}, this correlation implies a shift of the turnover
to lower frequencies while the AGN evolves and the jets expand. Here, we analyse
our sample in the context of this scenario.\par

The typical linear size of fully evolved RL AGN, i.e.\ FRI/FRII, is around
100\,kpc or higher~\citep{Pentericci2000}. Higher resolution data used in our
work show that the majority of IFRS are smaller than 100\,kpc as already mentioned by
\citet{GarnAlexander2008} and \citet{Middelberg2011}. However, generally, the
angular resolution of our observational data is not high enough to test whether
the correlation between intrinsic turnover frequency and linear size holds for
our sources. The plot from \citet{ODea1997} shows sources with linear sizes of
around 10\,kpc and smaller. Furthermore, the two correlated quantities are
redshift-dependent, resulting in an additional uncertainty for the
redshift-lacking IFRS sample.\par

Even for the most compact IFRS, the correlation from \citet{ODea1997} provides
only weak constraints. \citet{Middelberg2011} concluded that three IFRS (CS703,
ES427, ES509) are even smaller than $\mathrm{4.5\,kpc\times 2.1\,kpc}$ since
they do not show any evidence for being resolved at any of the five frequencies
used in their study. Following \citeauthor{ODea1997}, a source limited to that
linear size is expected to show a turnover at a rest-frame frequency of around
300\,MHz or higher. Already at a redshift of $z=1.5$, this turnover
would have been shifted out of the frequency range covered by our data.
This is consistent with our finding, that the radio SEDs of CS703 and
ES427 clearly follow single power laws down to observed frequencies of 100\,MHz
without indicating a turnover. In contrast, the radio SED of ES509 shows a
turnover at an observed frequency of around 130\,MHz. Assuming a rest-frame
peak frequency of at least 300\,MHz following from the \citeauthor{ODea1997}
correlation, this IFRS is expected to be at a redshift~$z\gtrsim 1.5$.\par

Going the other way around and using the fitted peak frequencies of
370\,MHz and 680\,MHz for ES1156 and CS164, the rest-frame peak
frequencies are above 1\,GHz and 2\,GHz, respectively, assuming that these sources are at $z\gtrsim 2$.
Following the correlation presented by \citet{ODea1997}, these objects are
expected to be smaller than $\sim 300$\,pc and $\sim 100$\,pc, respectively. In
all available maps, CS164 and ES1156 were found to be very compact.
In particular, CS164 was observed by \citet{Middelberg2011} at high angular
resolution and did not reveal any substructure, suggesting an angular size of
less than 1\,arcsec, consistent with the linear size estimated from the observed
turnover frequency.\par

\citet{Middelberg2011} find IFRS ES509 to be very compact and
suggest an angular size of less than $0.9\times 0.3\,\mathrm{arcsec}^2$,
corresponding to a physical size of $8\times 3\,\mathrm{kpc}^2$ for
$z\sim 1.5$. The \citet{ODea1997} correlation suggests an intrinsic
turnover frequency of 100\,MHz to 300\,MHz, consistent with the observed
turnover frequency of 130\,MHz for ES509 discussed above.\par

\subsection{Spectral components in the radio SED of IFRS}
\label{spectral_components}

The radio SED of an AGN may consist of one or more components. The core
component, mainly contributing at higher frequencies and potentially showing a
peak in the SED arising from SSA, is expected to represent the most recent
activity of the central engine and to be more time-variable than the steep
spectrum component of the extended emission. The latter---mainly given by the
lobes---represents the interaction between jet and the ambient medium and
therefore traces past activity of the core, with fading time scales of
$10^{4-5}$\,yrs~\citep{Miley2008}.
In contrast, the core component is expected to fade on timescales of the order
of $10^{1-3}$\,yrs~\citep{ODea1997,deVries2010}. Since the contributions of the
individual components to the total SED are changing with time because of these different
fading timescales, analysing the total radio SED for individual spectral components
provides a tool to study the activity status of the AGN. Since the core
components are likely to correspond to rest-frame frequencies of a few tens of
GHz, it is important that these high frequencies are considered when studying
radio sources even at low redshifts.\par

Our radio SED analysis showed that both components are found in the SEDs
of IFRS as presented in Sects.~\ref{powerlaw}, \ref{upturn_in_radiospectrum},
and \ref{spectra_turnover}. For example, it might be suggested that the core of
CS603 has been active more recently than that of e.g.\ CS703 because of the
additional---putatively inverted---spectral component at high frequencies. In
contrast, the core of CS164 might still be active since the core component is
dominating the total radio SED which is that of a GPS source. Our finding that
seven out of 34~IFRS in our sample are showing evidence for a core
component---all sources discussed in Sects.~\ref{upturn_in_radiospectrum} and
\ref{spectra_turnover}---is consistent with the fraction of IFRS with a detected
VLBI core. In addition, the radio SED of IFRS ES749 points at an additional
radio component at an observed frequency of around 200\,MHz. A similar behaviour
might be found in the radio SEDs of ES1018 and ES1259, but the frequency
coverage is not dense enough to substantiate this hypothesis. Importantly, a
source found to be consistent with a single power law up to 20\,GHz like CS703
might still have a flat or peaked spectrum component at higher frequencies.\par

\subsection{Radio spectral index as a function of IR and radio properties}
\label{alpha_asfunctionof_IRandRadio}

So far, we have focused on the radio SEDs of our IFRS sample and found that
the population of IFRS does not necessarily deviate from the general
RL source population with respect to the fraction of GPS and CSS
sources. However, IFRS were found to have steeper radio SEDs than the
ordinary AGN and star forming galaxy populations. Here, we link the radio SEDs
to the IR properties of IFRS.\par

\begin{figure*}
	\centering
		\includegraphics[width=0.495\hsize]{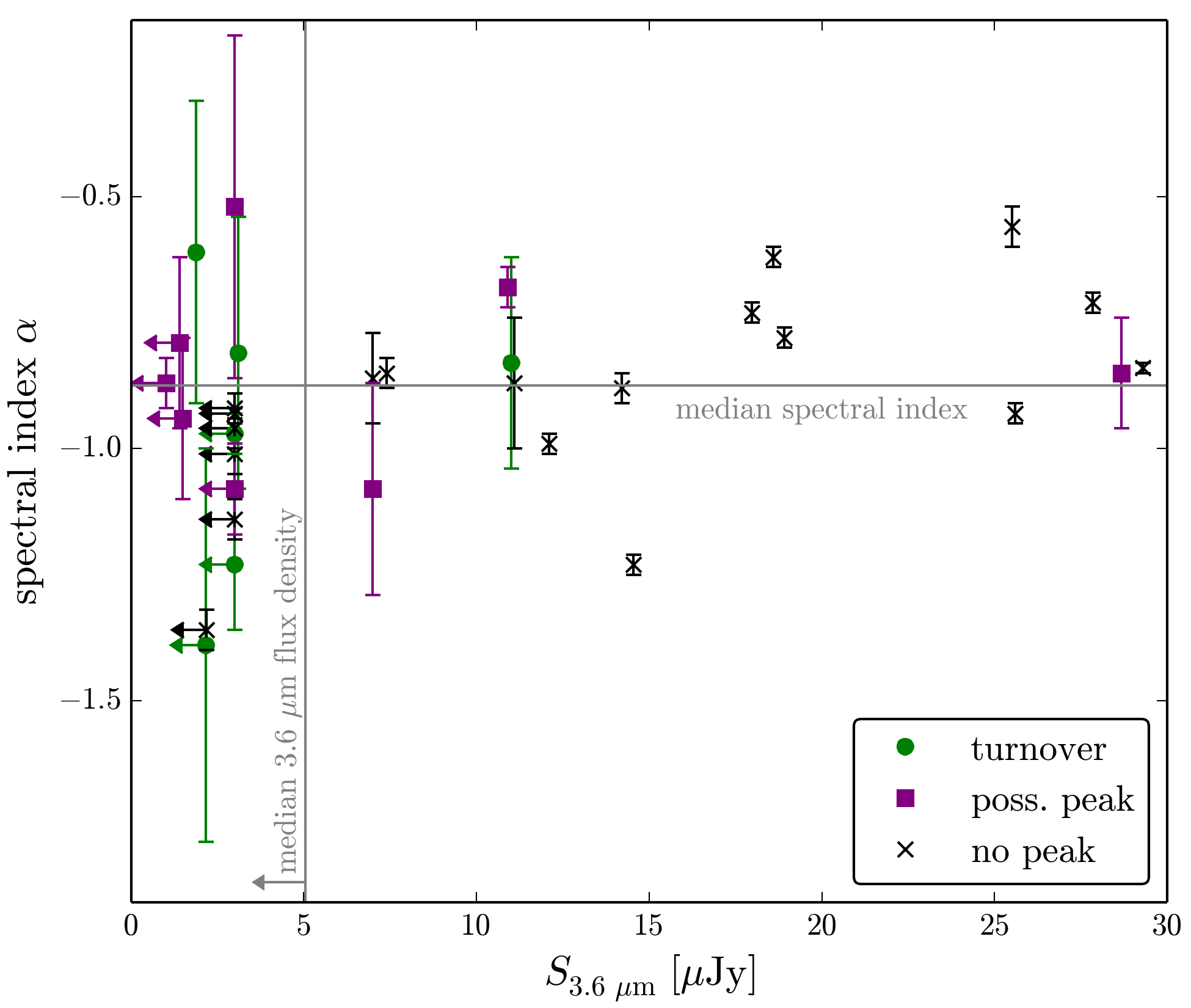}
		\includegraphics[width=0.495\hsize]{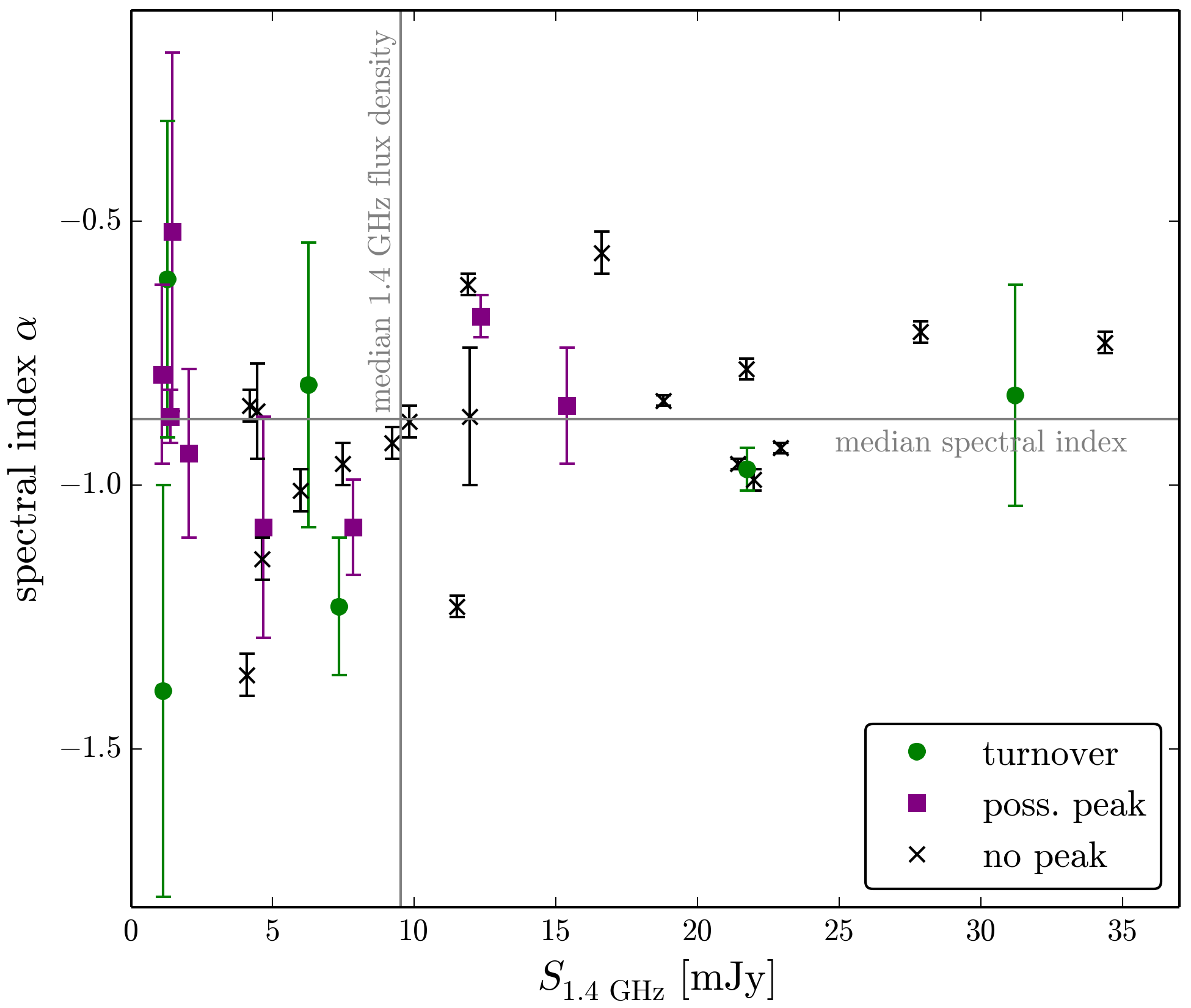}
		\includegraphics[width=0.495\hsize]{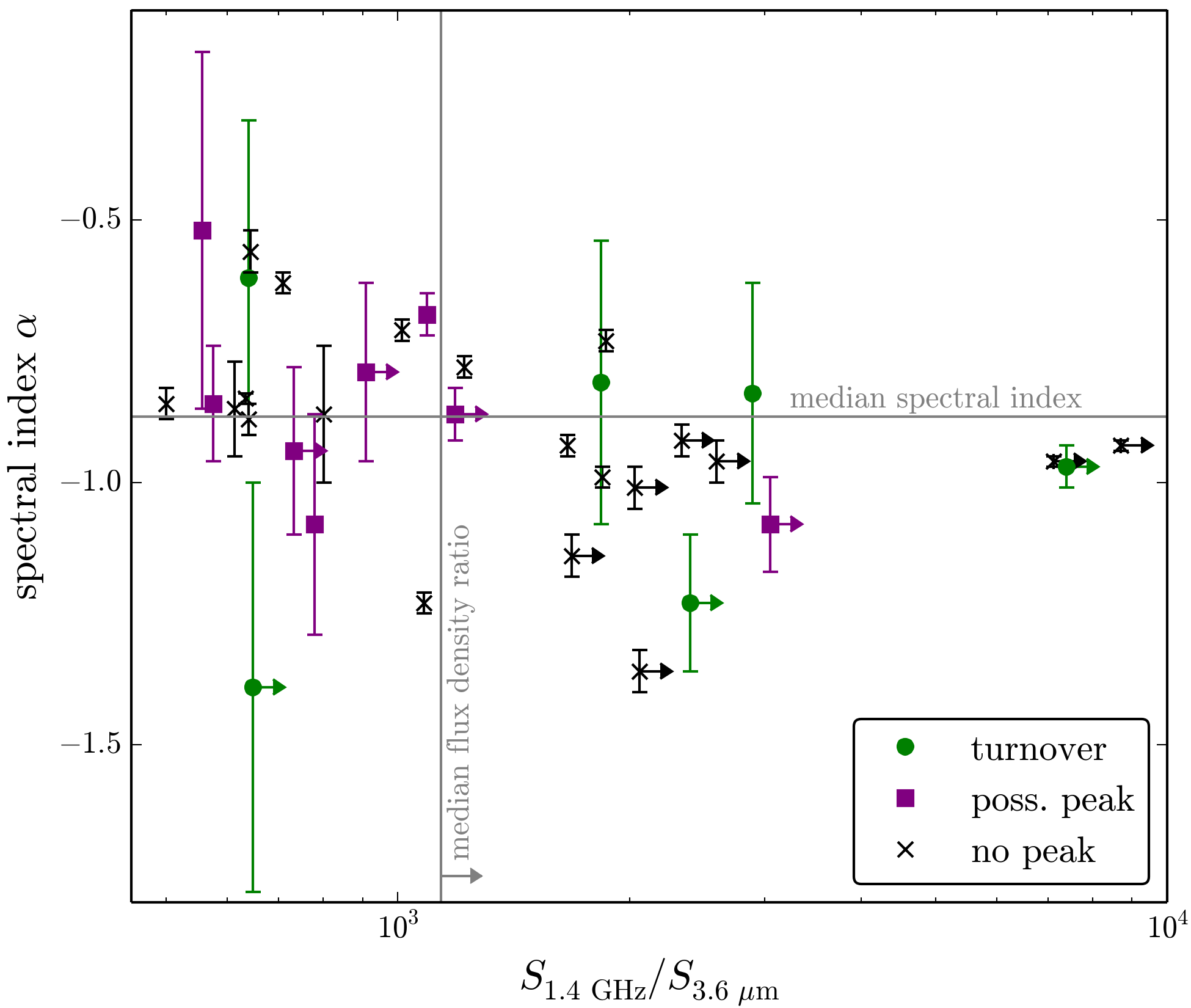}
		\includegraphics[width=0.495\hsize]{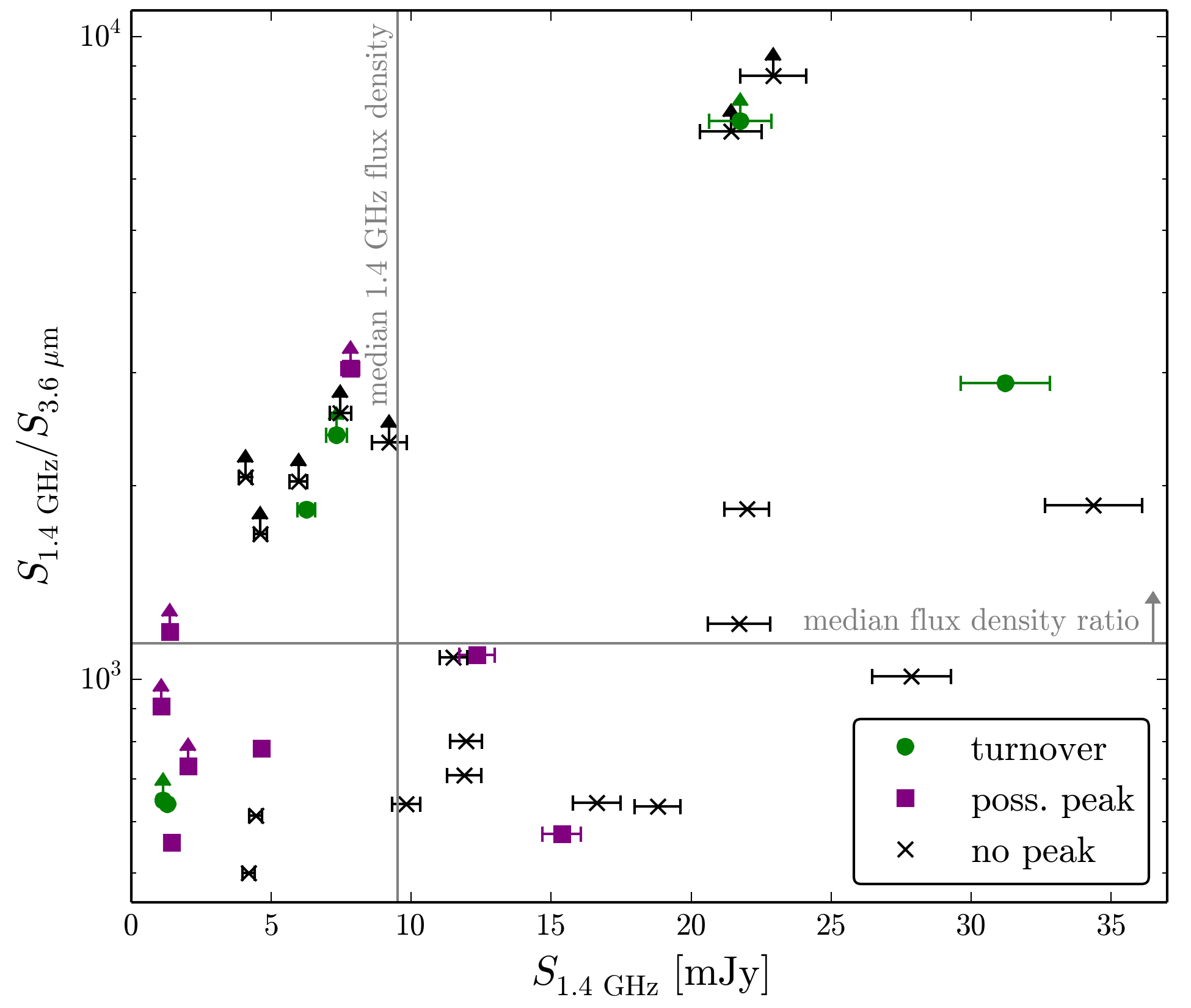}
		\caption{Radio spectral index~$\alpha$ as a function of the
		3.6\,$\mu$m flux density (top left), $\alpha$ as a function of the radio flux
		density~$S_\mathrm{1.4\,GHz}$ (top right), $\alpha$ as a function of
		the radio-to-IR flux density ratio~$S_\mathrm{1.4\,GHz}/S_\mathrm{3.6\,\mu m}$
		(bottom left), and $S_\mathrm{1.4\,GHz}/S_\mathrm{3.6\,\mu m}$ as a function
		of $S_\mathrm{1.4\,GHz}$ (bottom right) for our sample of 34~IFRS.  The
		symbols distinguish subsamples of different turnover properties. Green dots
		represent sources with a detected turnover. Sources with a possible peak are
		shown by purple squares, and sources without a peak by black crosses. The
		classification is described in Sect.~\ref{building_fitting_radioSEDs} and
		summarised for individual IFRS in Table~\ref{tab:sample}.
		Upper and lower limits are represented by arrows. The grey horizontal
		and vertical lines indicate the medians or limits of the median.}
	\label{fig:alpha_R_S1p4GHz_S3p6um}
\end{figure*}

As discussed in Sect.~\ref{steepening_spectra}, higher-redshift radio galaxies
have generally steeper radio SEDs. Connecting this correlation to the suggested
relation between 3.6\,$\mu$m flux density and redshift for
IFRS~\citep{Norris2011,Collier2014,Herzog2014}, IR-fainter IFRS would be
expected to have steeper radio SEDs. We tested our data for this potential
correlation and show the spectral index as a function of 3.6\,$\mu$m flux
density in Fig.~\ref{fig:alpha_R_S1p4GHz_S3p6um} (top left).
Our data provide statistically significant evidence for a correlation
between IR flux density and spectral index. When splitting our sample at a
3.6\,$\mu$m flux density of 15\,$\mu$Jy, we found only one~($13^{+20}_{-4}$\%)
out of eight IFRS in the IR-brighter subsample with a radio SED steeper than the
median, whereas 15~($58^{+9}_{-10}$\%) out of 26~IFRS in the IR-fainter
subsample have a steeper spectral index than the median. We also tested the data
based on a Spearman rank correlation test~(e.g.\ \citealp{Wall2012}) and found a
correlation coefficient~$r$ between 0.29 and 0.61, considering the
unknown relation between the IR-undetected sources. A coefficient of 1~($-1$)
corresponds to an ideal (anti-)correlation, whereas 0 indicates a lacking
correlation. The probability~$p$ that 3.6\,$\mu$m flux density and spectral
index are uncorrelated is between 0.046 and 0.00005. Therefore, we
reject the hypothesis that 3.6\,$\mu$m flux density and radio spectral index are
uncorrelated at a 0.05~significance level. For the control sample, we found a
correlation coefficient of $0.14$, indicating no evidence of a strong
correlation.\par

We also looked at the spectral index as a function of 1.4\,GHz flux density
as shown in Fig.~\ref{fig:alpha_R_S1p4GHz_S3p6um}~(top right) and found
no correlation, neither in the IFRS nor in the control sample. Based on
15.7\,GHz data, \citet{Franzen2014} showed that the general radio source
population with 15.7\,GHz flux densities above $\sim 25$\,mJy and below $\sim
1$\,mJy is dominated by flat ($\alpha >-0.5$) spectrum sources, whereas the
intermediate flux density range is dominated by sources with steep ($\alpha <
-0.5$) radio SEDs. We did not find any dependence at 1.4\,GHz. However, our
sample has been selected at lower frequencies and our sample covers only one
order of magnitude in 1.4\,GHz flux density, in contrast to the sample by
\citeauthor{Franzen2014}\par

The bottom left plot of Fig.~\ref{fig:alpha_R_S1p4GHz_S3p6um} shows
the spectral index~$\alpha$ as a function of the radio-to-IR flux density
ratio~$S_\mathrm{1.4\,GHz}/S_\mathrm{3.6\,\mu m}$. We find a Spearman rank
correlation coefficient between $-0.20$ and $-0.63$ and a probability between
0.12 and $3\times 10^{-5}$ that both quantities are uncorrelated, pointing at a
potential anti-correlation between spectral index and radio-to-IR flux density
ratio.\par

\subsection{Comparison of peaking and non-peaking sources}
\label{comparison_peak_nopeak}
As presented above, we found six IFRS that show a turnover in their
radio SEDs and 19~IFRS unambiguously without a peak in the frequency
range covered by our data. These subsamples are clearly marked in
Fig.~\ref{fig:alpha_R_S1p4GHz_S3p6um}, showing the IFRS sample in the parameter
spaces of spectral index, radio flux density, IR flux density, and radio-to-IR
flux density ratio. We find that five out of the six peaking IFRS have
3.6\,$\mu$m flux densities of 3\,$\mu$Jy and lower, below the median of the
entire IFRS sample. In contrast, these six peaking IFRS have 1.4\,GHz flux
densities from the entire flux density range covered by the total IFRS sample.
Similarly, the radio-to-IR flux density ratios of the IFRS with a turnover in
their radio SED do not differ from those of the non-peaking subsample.\par

Our subsample of peaking sources putatively contains GPS and young CSS sources,
whereas the non-peaking subsample is expected to contain older CSS sources
peaking at observed frequencies below 150\,MHz. Since we found the peaking
subsample to be IR-fainter, we expect these IFRS to be at higher redshifts,
following the suggested correlation between 3.6\,$\mu$m flux density and
redshift for IFRS~\citep{Norris2011,Collier2014,Herzog2014}. On the other hand,
the IR faintness and expected higher redshifts of the peaking subsample make a
detection of the peak less likely since the turnover would be redshifted out of
the frequency regime covered by our data. Redshift information is crucial to
disentangle these two effects---evolution and cosmology---, contributing in
opposite directions.\par

\subsection{Radio SED of IFRS CS618}
\label{S618}

\begin{figure}
	\centering
		\includegraphics[width=\hsize]{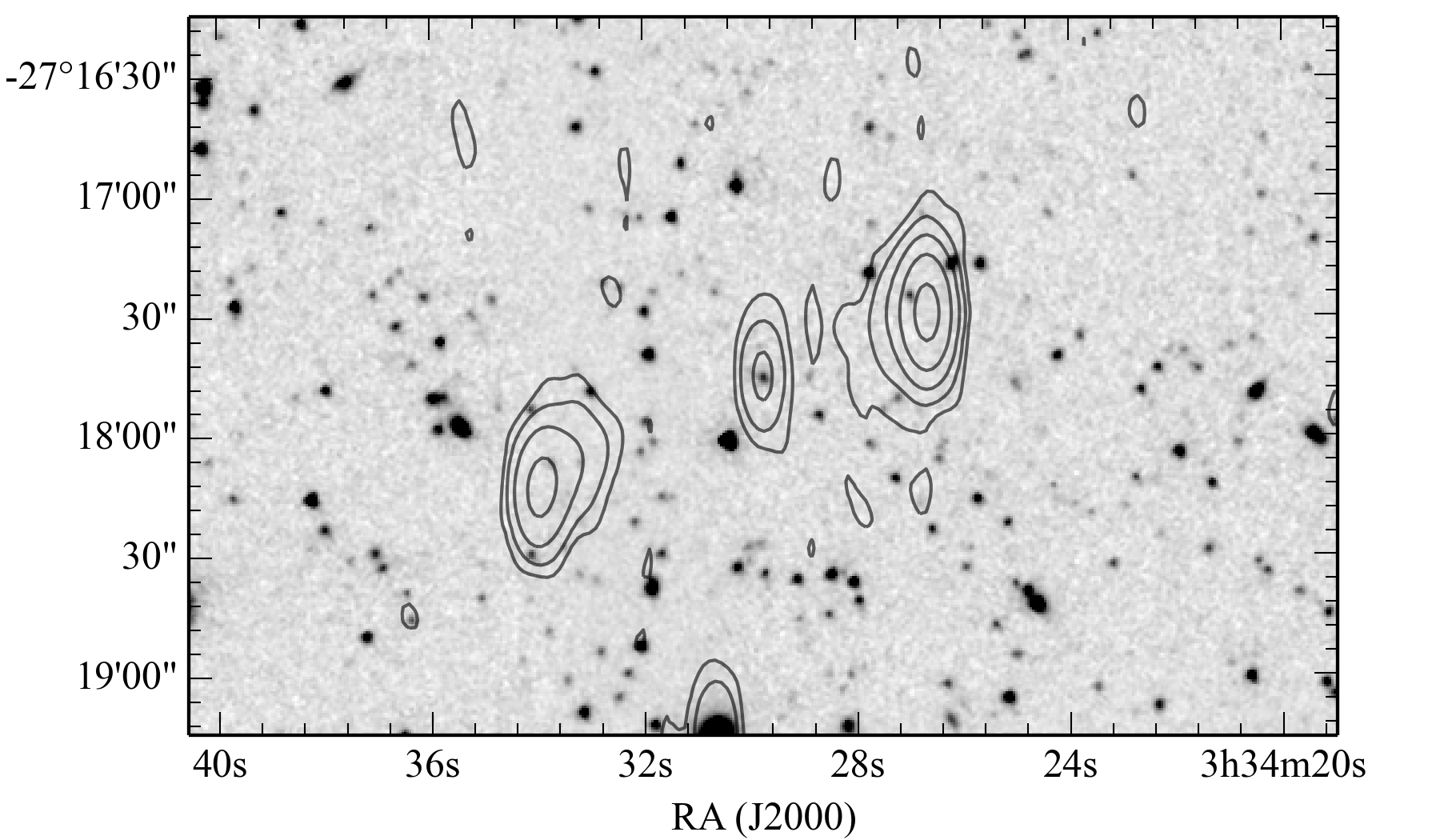}
		\caption{3.6\,$\mu$m SWIRE map (\citealp{Lonsdale2003}; greyscale) of CS618
		overplotted by the 1.4\,GHz ATLAS DR3~\citep{Franzen2015} contours.
		Contours start at $3\sigma$ and increase by factors of 4. CS618 is an evolved
		FRII radio galaxy with an unresolved core (centre of the shown map) and two
		slightly resolved lobes.}
	\label{fig:S618}
\end{figure}

The IFRS~CS618 is peculiar and has a different morphology than any other
source in our IFRS sample. The 3.6\,$\mu$m map of CS618 overplotted by the
1.4\,GHz ATLAS DR3 contours is shown in Fig.~\ref{fig:S618}.
It appears as a double-lobed FRII radio galaxy with an angular size of around
1.7\,arcmin. Its total radio SED (containing the core and the two lobes) is well
described by a single power law with a spectral index $\alpha = -1.06 \pm 0.05$
(see Fig.~\ref{fig:radioSEDs}). All three components are separately detected at
325\,MHz, 610\,MHz, 1.4\,GHz and 2.3\,GHz, but are indistinguishable at
200\,MHz, 843\,MHz, and 844\,MHz because of the lower resolution. The radio SEDs
of all three individual components are well described by single power laws. The
central component (1.6\,mJy at 1.4\,GHz) is unresolved with a spectral index
between 325\,MHz and 2.3\,GHz of $+0.33 \pm 0.17$, whereas the two lobes are
slightly resolved with 1.4\,GHz flux densities of 34.5\,mJy and 11.4\,mJy, and
spectral indices between 150\,MHz and 2.3\,GHz of $-1.01 \pm 0.09$ and $-1.23
\pm 0.13$. These characteristics are common for FRII radio galaxies~(e.g.\
\citealp{Hovatta2014}) and so CS618 seems to differ from the majority of IFRS in
our sample, which are dominated by more compact and putatively younger radio
sources as discussed above.\par

\subsection{Comparison to sources used in broad-band SED modellings}
\label{individual_sources}

\citet{Herzog2014} presented the first redshift-based SED modelling of IFRS and
find all three studied IFRS in agreement with scaled templates of 3C\,48 and
3C\,273. Two of these three IFRS are also in our sample: CS265 and CS713. The
radio SEDs of these IFRS were found to be in good agreement with single
power laws with spectral indices of $-0.84 \pm 0.01$ and $-0.56 \pm 0.04$,
respectively. 3C\,48 is a CSS source with a spectral index of $\alpha
\sim -0.8$, and the core of the RL quasar~3C\,273 is known to have a flat radio SED.
Based on the spectral index found in our work, the core of 3C\,273 seems to
provide an inadequate template to explain the characteristics of IFRS~CS265. In
contrast, the spectral index of CS713 matches the flat radio SED known for
3C\,273, whereas the steeper SED of 3C\,48 disagrees with the characteristics of
CS713.\par

An extended sample of SED templates was used by \citet{Herzog2015b} to
constrain the broad-band SED of six IFRS observed with \textit{Herschel}.
All sources are also in our radio sample of IFRS and we found radio spectral
indices between $-0.92$ and $-1.08$, i.e.\ all have steep or ultra-steep SEDs.
\citeauthor{Herzog2015b} find only the redshifted broad-band SED
templates of the spiderweb galaxy, Cygnus~A, 3C\,48, and 3C\,273 to be in
agreement with the data, even if the templates were scaled in luminosity and extinction
was added. The flat radio SED of the core of 3C\,273 is in clear disagreement
with the steep radio SEDs found for all six sources. Also the radio SED of
3C\,48 is flatter than any radio SED in that sample of six IFRS, although an
overlap with the flattest sources in the IFRS sample cannot be ruled out. In
contrast, both the spiderweb galaxy and Cygnus~A are known to have ultra-steep
radio SEDs with spectral indices of around $-1.3$ and $-1.2$, respectively.
Thus, the radio spectral indices measured in our work add evidence to the
conclusion presented by \citeauthor{Herzog2015b} that the spiderweb galaxy
and Cygnus~A provide adequate templates to explain the broad-band
characteristics of IFRS.\par

\section{The radio and multi-wavelength SED of IFRS xFLS~478}
\label{IFRS_xFLS478}
In Sect.~\ref{discussion_radioSEDs}, we analysed the broad radio SEDs of a large
sample of IFRS in the ELAIS-S1 and CDFS fields. To complement this study, we
present the detailed analysis of one IFRS (xFLS\,478) in the following. This
study is not limited to the radio regime, but also links to the far-IR~(FIR)
detections of this source. IFRS xFLS\,478 is particularly suitable for this study since it
provides the highest-frequency radio data point of an IFRS (at 105\,GHz as
presented in Sect.~\ref{PdBI_calibration}) and ancillary data in the radio and IR
regime, described in Sect.~\ref{ancillarydata_xFLS478}.\par

\begin{figure}
	\centering
		\includegraphics[width=\hsize]{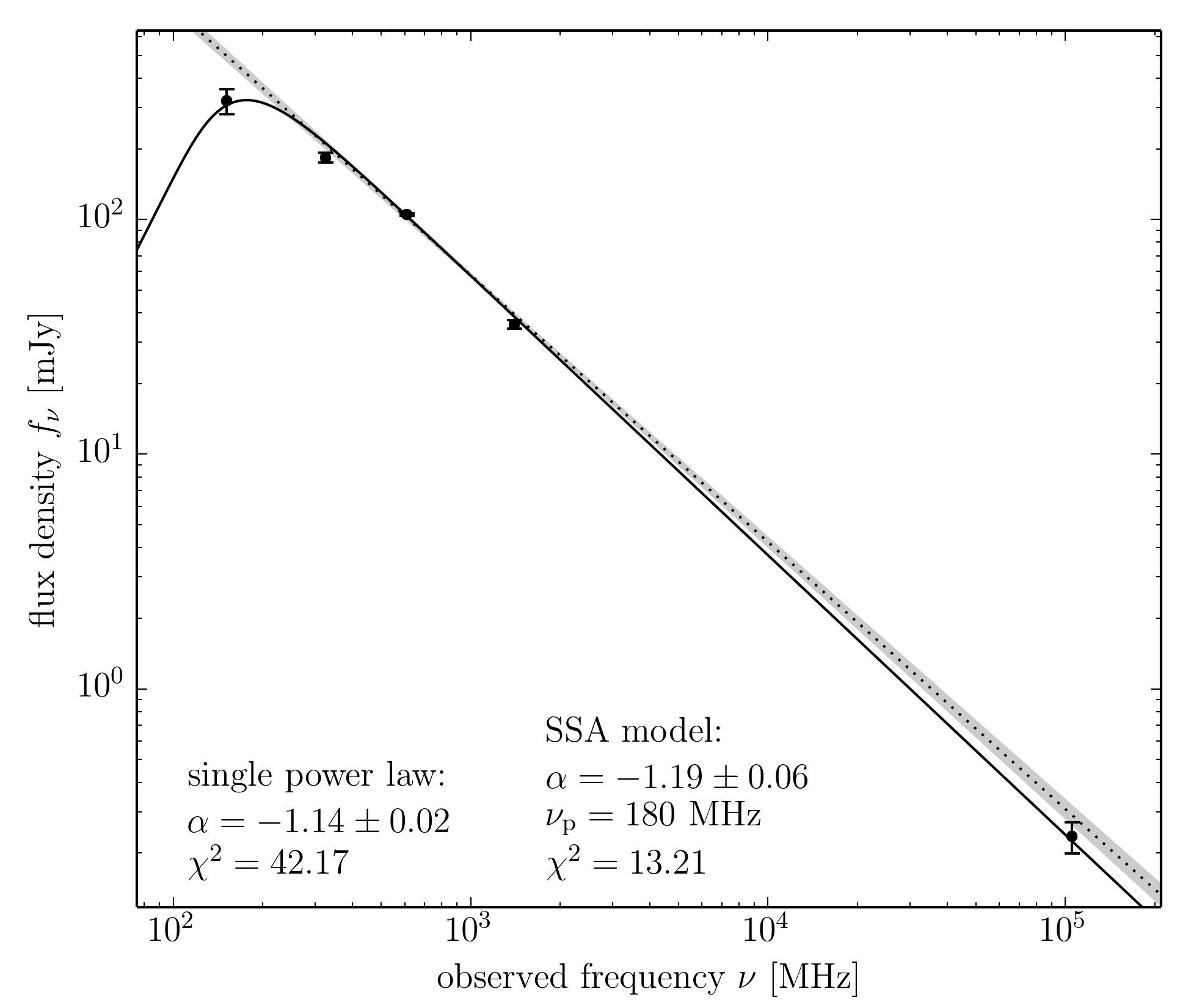}
		\caption{Radio SED of IFRS xFLS\,478. Error bars represent $1\sigma$
		uncertainties. The black solid line shows the best fit of an SSA model. The
		fitted single power law is shown as black dotted line.
		The shaded area represents the $1\sigma$ uncertainty of the single power
		law fit.}
	\label{fig:radioSED_xFLS478}
\end{figure}

Figure~\ref{fig:radioSED_xFLS478} shows the radio SED of this source---including
all available data between 150\,MHz and 105\,GHz---, similar to the SEDs shown
in Fig.~\ref{fig:radioSEDs}.
To describe the radio emission over this wide frequency range of almost three
orders of magnitude, we followed the same approach as described in
Sect.~\ref{building_fitting_radioSEDs}. Fitting a single power law to the data resulted in
a spectral index $\alpha = -1.14\pm 0.02$. However, the low-frequency data at
150\,MHz and 325\,MHz depart from the single power law, indicating a
potential turnover. An SSA model was found to better describe the data,
resulting in a turnover at 180\,MHz and a spectral index of $-1.19 \pm 0.06 $ at
the high-frequency side of the synchrotron bump. However, a turnover can only be
suggested based on the available data; its frequency remains unclear. In any
case, IFRS xFLS\,478 has a steep radio SED and might---depending on the
definition as discussed in Sect.~\ref{ultrasteep_spectra}---be classified as a
USS source according to both fits.\par

If the radio SED of xFLS\,478 indeed follows the fitted single power law or the
SSA model up to an observed frequency above 100\,GHz, this provides interesting
new constraints on the properties of IFRS. As shown by \citet{Collier2014} and
\citet{Herzog2014}, all known redshifts of IFRS are in the range $1.7\lesssim z
\lesssim 3.0$. Assuming that xFLS\,478 is at a similar redshift, the observed
105\,GHz detection is at a rest-frame frequency $\sim 300$\,GHz. It is known
that the thermal free-free and dust emission in star forming galaxies start to
dominate over non-thermal synchrotron emission at around 100\,GHz~(e.g.\
\citealp{Murphy2009}) as discussed in Sect.~\ref{upturn_in_radiospectrum}. This
would imply for xFLS\,478 that the AGN emission of this source dominates over
the emission from star forming activity even in the mm~regime despite the
\textit{Herschel} detection at 250\,$\mu$m. The alternative explanation for a
flattening of the radio SED at higher frequencies---a dominating AGN core as
discussed in Sects.~\ref{upturn_in_radiospectrum} and
\ref{spectral_components}---can also be excluded for xFLS\,478 up to this
frequency. However, an additional, flat radio SED component might still start to
dominate at higher frequencies as discussed above.\par

The same behaviour---following a power law up to the mm regime---was found for
the source F00183-7111~\citep{Norris2012}, referred to as 00183. 00183 is one of the
most luminous ULIRGs, being heavily obscured and undergoing vigorous star
forming activity, at $z=0.3276$. In VLBI observations, a core-jet structure was
found in its centre with an extension of only 1.7\,kpc, however the source is
radio-loud. \citeauthor{Norris2012} suggested that this source is undergoing a
transition from a merging starburst with a quasar-mode AGN in its centre to an
RL quasar. \citet{Emonts2011a} and \citet{Emonts2011b} found three HzRGs to
follow single power laws up to 36\,GHz (115\,GHz in the restframe).\par

We emphasise that while the suggested single power law or SSA model for
xFLS\,478 are the simplest explanation, it is based on only one detection in the
mm regime. Other interpretations are also possible, such as a steepening SED
above 1.4\,GHz, followed by a minimum between 10\,GHz and 100\,GHz, and
increasing thermal dust emission---dominating over synchrotron emission---at
105\,GHz. Further high-frequency observations would be required to distinguish
between these hypotheses In Sect.~\ref{upturn_in_radiospectrum}, we presented
one IFRS that did not follow a power law up to high frequencies but showed
deviations at observed frequencies around 20\,GHz. Most likely, this deviation
is caused by the flat or inverted SED of an AGN core.\par

We now consider the multi-wavelength SED of IFRS xFLS\,478, including all
ancillary data presented in Sect.~\ref{ancillarydata_xFLS478} and following the
approach used by \citet{Herzog2015b}. In this modelling, SED templates of
different galaxy classes typically found at high redshifts---including star
forming galaxies with and without AGN, Seyfert galaxies, and RL AGN---were
used, shifted in the redshift range $0.5\leq z \leq 12$, scaled in luminosity,
and obscured by additional dust. These templates were then tested whether they
are consistent with the photometric data of IFRS. In this comprehensive
approach, we did not find any SED template that could reproduce the
characteristics of IFRS xFLS\,478 in the redshift regime $0.5\leq z \leq 12$. In
particular, the most promising templates---RL AGN at high redshifts,
e.g.\ HzRGs---were found to be inconsistent with the FIR detections of
xFLS\,478. \citet{Herzog2015b} observed six IFRS with \textit{Herschel}
that are only slightly radio-fainter than xFLS\,478 but none was detected; the
detection sensitivities were similar to the observed FIR flux densities of
xFLS\,478.\par

\begin{figure}
	\centering
		\includegraphics[width=\hsize]{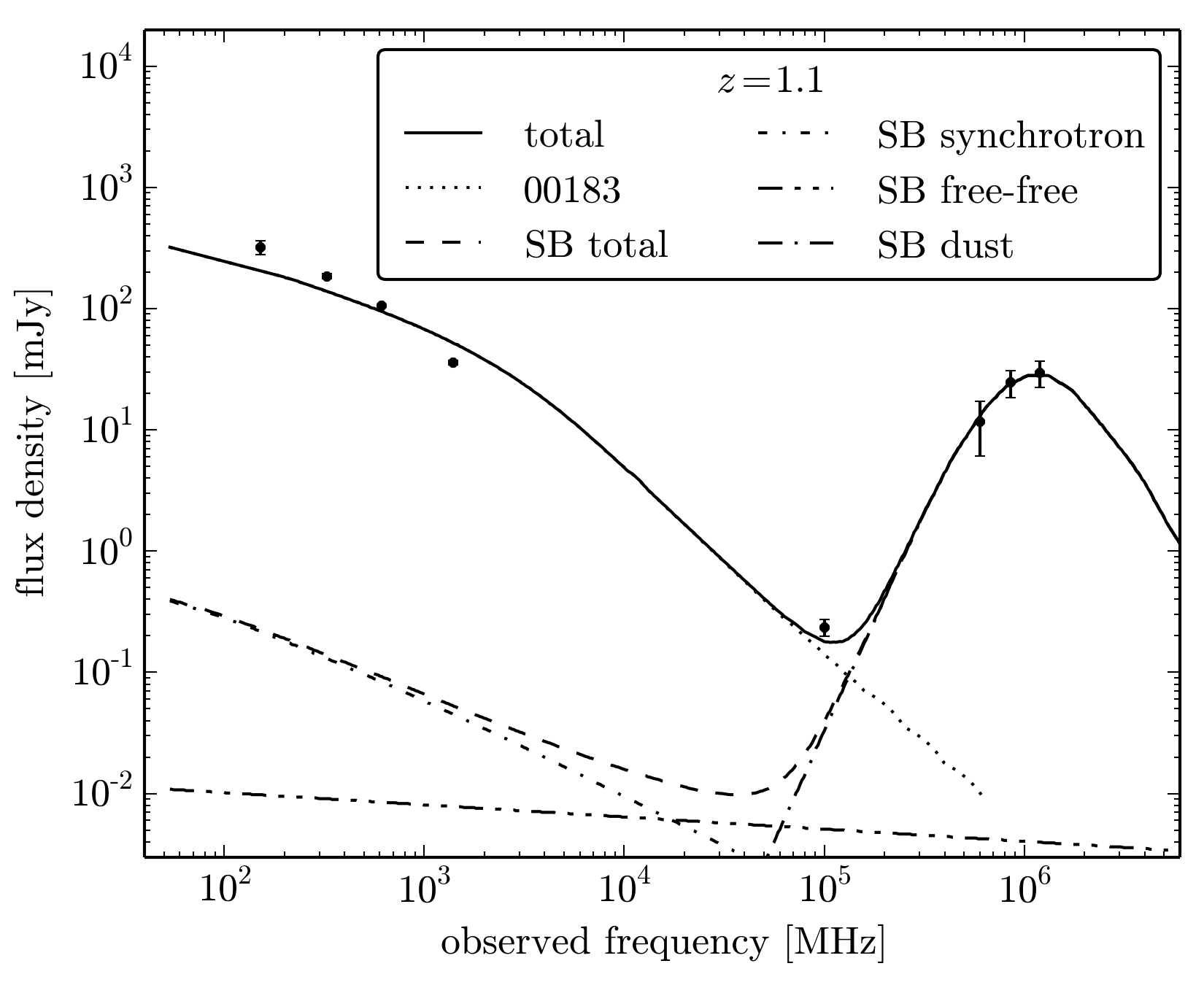}
		\caption{Modelling the radio-FIR SED of IFRS~xFLS\,478. The total SED
		(solid line) is composed of the SED of a star forming galaxy (dashed line)
		from \citet{Murphy2009} and the radio SED of 00183 (dotted line)
		from \citet{Norris2012}. The star forming SED template is again composed of
		synchrotron, free-free, and thermal dust components. The best model was found
		at $z=1.1$ and is shown in this figure. Black dots represent the photometric
		data points of xFLS\,478 with respective $1\sigma$ uncertainties.}
	\label{fig:xFLS478_radioFIRmodel}
\end{figure}

Consequently, we asked the question what properties a galaxy would need to have
to be consistent with the photometric constraints of xFLS\,478. To account for
the HerMES FIR detections with flux densities of a few tens of mJy, we used
the radio-FIR SED template of a star forming galaxy from \citet{Murphy2009}.
This template is composed of synchrotron, free-free, and thermal dust
components as shown in Fig.~\ref{fig:xFLS478_radioFIRmodel} for $z=1.1$.
However, the observed radio emission of xFLS\,478 cannot be explained by star foming
activity at any redshift. Therefore, we added the radio emission from an RL~AGN,
consistent with the finding that the majority of IFRS---if not all---contain
AGN~\citep{Herzog2015a}. We used the radio SED of 00183 from
\citet{Norris2012}. This source is known to host an RL AGN and to show similarities to xFLS\,478 as discussed
above. We shifted these templates in the redshift range $0.5\leq z \leq 8.0$ and built the
total SED by summing both individual templates, each of them scaled by an
individual factor. The best modelling was found at $z=1.1$ and is shown in
Fig.~\ref{fig:xFLS478_radioFIRmodel}.
We note that the fractional contributions from synchrotron, free-free, and
thermal dust emission to the star forming galaxy template were fixed as
described by \citeauthor{Murphy2009} and were not varied in our modelling.\par

The photometric data of the IFRS are well described by the model at $z=1.1$,
particularly in the FIR and mm~regimes. The radio SED of xFLS\,478 seems to be
flatter at higher frequencies than that of 00183. \citet{Norris2012} fitted a
broken power law to the radio SED of 00183, finding a break at around 5\,GHz and
spectral indices of $-0.43$ and $-1.49$ at low and high frequencies,
respectively. We measured a spectral index of $-1.19$ for xFLS\,478 as described
above and did not find clear evidence for a break at higher frequencies.\par

At $z=1.1$, the 00183 radio template had to be scaled up in luminosity by a
factor of 3.0. The star forming galaxy template had to be scaled up by a factor
of 1.9, giving a star formation rate of around 170\,M$_\odot$\,yr$^{-1}$. This
model, if confirmed, makes xFLS\,478 the lowest-redshift IFRS known.\par

\section{Conclusion}
\label{conclusion}

We built radio SEDs for 34~IFRS in the CDFS and ELAIS-S1 fields, covering the
frequency range between 150\,MHz and 34\,GHz. Based on these SEDs, we found the
vast majority of IFRS ($74^{+6}_{-9}$\%) to show steep radio SEDs defined by
$\alpha < -0.8$. $6^{+7}_{-2}$\% of the IFRS in our sample are classified as USS
sources~($\alpha < -1.3$) and are therefore good candidates for high-redshift
sources. The sample of IFRS shows statistically significantly steeper radio SEDs
than the broader RL~AGN population. The median spectral index in our IFRS sample
is $-0.88$. One IFRS was found to show a flattening or upturning radio
SED at 20\,GHz, indicating an additional core component and emphasising the
importance of considering high-frequency radio data when studying radio sources,
irrespective of their redshift. Our finding that IR-fainter IFRS have steeper
radio SEDs supports the hypothesis that IR-fainter IFRS are at higher
redshifts.\par

We found $3^{+6}_{-1}$\% of our sample are GPS sources and $\geq 15^{+8}_{-4}$\%
are CSS sources. These numbers are consistent with the general fraction of GPS and
CSS sources in the RL AGN population. This finding implies that at least some
IFRS are young AGN in the earliest stages of their evolution to powerful and
extended FRI/FRII radio galaxies. However, the intrinsic fraction of GPS and CSS
sources in the IFRS population might be higher than in the general RL
source population if IFRS are at higher redshifts. Generally, IFRS are
prototypical for the class of CSS sources because of their steep radio SEDs and
their compactness. Our analysis showed that IFRS with an observed peak in their radio
SED are IR-fainter than IFRS without a turnover. However, we do not find
evidence that the radio flux densities or the radio-to-IR flux density ratios of
peaking IFRS differ from those of the non-peaking subsample.\par

We also carried out a detailed analysis of the broadband SED of
IFRS~xFLS\,478. This source was observed with the PdBI at 105\,GHz and provided
the highest-frequency radio detection of an IFRS. The source was found to have a
steep radio SED, potentially indicating a turnover at around 150\,MHz. We did not observe an
upturn or flattening in the radio SED at high frequencies, indicating that
synchrotron emission dominates over thermal dust emission at least down to a rest-frame
frequency of 300\,GHz (1\,mm) if the source is at $z\gtrsim 2$.\par

Modified SED templates of known galaxies were found to be inconsistent with the
multi-wavelength data of xFLS\,478. However, the data are well described by a
radio-FIR SED template composed of a star forming galaxy and an RL~AGN at
$z=1.1$ which would make this object the IFRS with the lowest known
redshift. This model suggests a star formation rate of around
170\,$M_\odot$\,yr$^{-1}$.\par

\begin{acknowledgements}
We thank our referee Alastair Edge for valuable suggestions which helped to
improve this paper.

We thank the IRAM staff for carrying out the PdBI observations and helping with
data calibration.

NS is the recipient of an ARC Future Fellowship. BE acknowledges funding through
the European Union FP7 IEF grant nr.~624351.

This research has made use of the NASA/IPAC Extragalactic Database (NED)
which is operated by the Jet Propulsion Laboratory, California Institute of
Technology, under contract with the National Aeronautics and Space
Administration.

This research has made use of data from HerMES project
(http://hermes.sussex.ac.uk/). HerMES is a Herschel Key Programme utilising
Guaranteed Time from the SPIRE instrument team, ESAC scientists and a mission
scientist.

The HerMES data was accessed through the Herschel Database in Marseille (HeDaM -
http://hedam.lam.fr) operated by CeSAM and hosted by the Laboratoire
d'Astrophysique de Marseille.

The Australian SKA Pathfinder is part of the Australia Telescope National
Facility which is funded by the Commonwealth of Australia for operation as a
National Facility managed by CSIRO. This scientific work uses data obtained from
the Murchison Radio-astronomy Observatory~(MRO), which is jointly funded by the
Commonwealth Government of Australia and State Government of Western Australia.
The MRO is managed by the CSIRO, who also provide operational support to ASKAP.
We acknowledge the Wajarri Yamatji people as the traditional owners of the
Observatory site. The work was supported by iVEC through the use of advanced
computing resources at the Pawsey Advanced Supercomputing Centre.”

Support for the operation of the MWA is provided by the Australian
Government~(NCRIS), under a contract to Curtin University administered by
Astronomy Australia Limited. We acknowledge the Pawsey Supercomputing Centre
which is supported by the Western Australian and Australian Governments.

\end{acknowledgements}

-------------------------------------------------------------------

\bibliographystyle{aa} 
\bibliography{references} 

\appendix

\section{Photometric data of IFRS in ELAIS-S1 and CDFS}
\label{datasection}

The photometric data used to study the radio SEDs of 15~IFRS in ELAIS-S1 and
19~IFRS in CDFS are summarised in Tables~\ref{tab:data_ELAIS} and
\ref{tab:data_CDFS}, respectively.

\begin{sidewaystable*}
\begin{landscape}
	\caption{Flux densities used in this work for the IFRS located in the ELAIS-S1
	field. $S_\mathrm{200\,MHz}$ denotes the GLEAM flux density in the deep image
	(60\,MHz bandwidth). The last column lists the ATLAS DR3 in-band spectral index
	between 1.40\,GHz and 1.71\,GHz.}
	\label{tab:data_ELAIS}
\centering
 \begin{tabular}{c c c c c c c c c c}
 \hline \hline
  IFRS   &  $S_\mathrm{200\,MHz}$& $S_\mathrm{610\,MHz}$ & $S_\mathrm{843\,MHz}$ & $S_\mathrm{1.4\,GHz}$ & $S_\mathrm{2.3\,GHz}$  & $S_\mathrm{4.8\,GHz}$ & $S_\mathrm{8.6\,GHz}$  & $S_\mathrm{34\,GHz}$ & $\alpha_{1.40}^{1.71}$ \\
  ID     & [mJy]                & [mJy]                & [mJy]                & [mJy]                 & $[\mathrm{mJy}]$      & [mJy]                & [mJy]                & [mJy]               &                       \\ \hline
  ES5    & $  143 \pm 8      $  & $26.6 \pm  6.7 $     & $22.5  \pm 2.0$      & $11.51\pm 0.49$      & $ 5.21  \pm 0.72$     & $ \dots       $      & $ \ldots       $     & $ \ldots         $  &$ \ldots               $\\
  ES66   & $  147 \pm 9      $  & $67.0 \pm  19.6$     & $61.6  \pm 4.5$      & $34.37\pm 1.74$      & $ 20.84 \pm 2.16$     & $ \ldots      $      & $ \ldots       $     & $ \ldots         $  &$ \ldots               $\\
  ES201  & $ 46 \pm 13       $  & $8.5  \pm  2.1 $     & $<11.3^\star   $      & $4.62\pm  0.23$      & $ 2.18  \pm 0.26$     & $ \ldots      $      & $ \ldots       $     & $ \ldots         $  &$-1.26   \pm     0.12  $ \\
  ES419  & $<21              $  & $2.2  \pm  0.6 $     & $<4.3^\star    $      & $1.45\pm  0.09$      & $ <1.31^\star    $     & $<0.39        $      & $<0.33         $     & $ \ldots         $ &$-1.07   \pm     0.45   $\\
  ES427  & $  119 \pm 8      $  & $47.7 \pm  11.9$     & $36.8  \pm 2.1$      & $21.42\pm 1.10$      & $ 12.63 \pm 1.26$     & $6.54 \pm 0.46$      & $2.83  \pm 0.36$     & $0.575  \pm 0.098$  &$-0.95  \pm      0.10   $\\
  ES509  & $  120 \pm 8      $  & $49.2 \pm  12.3$     & $36.6  \pm 2.2$      & $21.75\pm 1.11$      & $ 13.18 \pm 1.33$     & $5.94 \pm 0.45$      & $3.51  \pm 0.39$     & $0.550  \pm 0.096$  &$-0.99 \pm       0.10   $\\
  ES645  & $<20              $  & $10.7 \pm  3.1 $     & $8.5   \pm 2.7$      & $4.66 \pm 0.25$      & $ 2.67  \pm 0.41$     & $ \ldots      $      & $ \ldots       $     & $ \ldots         $  &$0.05  \pm        0.33  $\\
  ES749  & $  60 \pm 15      $  & $15.7 \pm  4.7 $     & $12.6  \pm 1.7$      & $9.22 \pm 0.62$      & $ 4.99  \pm 0.69$     & $2.63 \pm 0.29$      & $0.82  \pm 0.26$     & $ \ldots         $  &$ \ldots                $\\
  ES798  & $   60 \pm 11     $  & $14.6 \pm  3.7 $     & $12.7  \pm 1.7$      & $7.48 \pm 0.38$      & $ 4.37  \pm 0.51$     & $3.07 \pm 0.37$      & $<0.33         $     & $ \ldots         $  &$-0.70 \pm        0.19  $\\
  ES973  & $ <97^\star        $  & $12.1 \pm  3.0 $     & $17.2  \pm 2.6$      & $7.83 \pm 0.32$      & $ 4.05  \pm 0.44$     & $1.93 \pm 0.29$      & $<0.27         $     & $<0.385          $  &$-0.84 \pm        0.13  $\\
  ES1018 & $  113 \pm 9      $  & $58.0 \pm  16.9$     & $42.0  \pm 2.5$      & $27.87\pm 1.40$      & $ 18.89 \pm 1.92$     & $ \ldots      $      & $ \ldots       $     & $ \ldots         $  &$-0.74 \pm        0.13  $\\
  ES1021 & $ <131^\star       $  & $30.3 \pm  7.6 $     & $24.8  \pm 1.6$      & $15.38\pm 0.69$      & $ 10.88 \pm 1.09$     & $ \ldots      $      & $ \ldots       $     & $ \ldots         $  &$-0.76 \pm        0.10  $\\
  ES1156 & $  35 \pm 7       $  & $55.1 \pm  13.8$     & $46.6  \pm 2.9$      & $31.22\pm 1.60$      & $ 20.13 \pm 2.02$     & $ \ldots      $      & $ \ldots       $     & $ \ldots         $  &$-0.81 \pm        0.10  $\\
  ES1239 & $  111 \pm 8      $  & $48.9 \pm  14.3$     & $37.1  \pm 2.3$      & $21.72\pm 1.11$      & $ 15.31 \pm 1.54$     & $ \ldots      $      & $ \ldots       $     & $ \ldots         $  &$-0.83 \pm        0.10  $\\
  ES1259 & $   67 \pm 10     $  & $  \ldots      $     & $ \mathrm{af} $      & $4.08 \pm 0.22$      & $ 2.31  \pm 0.82$     & $ \ldots      $      & $ \ldots       $     & $ \ldots         $  &$\ldots                 $\\ \hline
 \end{tabular}
\tablefoot{Flux density upper limits resulting from confusion are marked by ($^\star$). Sources for which the flux density could not be measured at the respective frequency because of image artefacts are marked by ``af''. Sources that were outside the survey fields or not targeted by the observations at the respective frequency are represented by ellipsis dots~($\ldots$).}
\end{landscape}
\end{sidewaystable*}

\begin{table*}
	\caption{Flux densities used in this work for the IFRS located in the CDFS. $S_\mathrm{200\,MHz}$ denotes the GLEAM flux density in the deep image
	(60\,MHz bandwidth). The last column lists the ATLAS DR3 in-band spectral index between 1.40\,GHz and 1.71\,GHz.}
	\label{tab:data_CDFS}
\centering
 \begin{tabular}{c c c c c c c c c}
 \hline \hline
 IFRS   & $S_\mathrm{150\,MHz}$ & $S_\mathrm{200\,MHz}$ & $S_\mathrm{325\,MHz}$ & $S_\mathrm{610\,MHz}$&  $S_\mathrm{843\,MHz}$ & $S_\mathrm{844\,MHz}$ & $S_\mathrm{1.4\,GHz}$ & $S_\mathrm{2.3\,GHz}$\\
 ID     & [mJy]                & [mJy]                & [mJy]                & [mJy]               & [mJy]                 & $\mathrm{[mJy]}$     & [mJy]                & $\mathrm{[mJy]}$    \\ \hline
 CS94   & $ 89    \pm 27   $   &  $<104^\star       $  & $  \mathrm{af}     $ & $     \ldots     $  & $ <55.6           $   & $ <27.7^\star   $     & $ 11.96 \pm   0.57$ & $ 8.12   \pm 0.93  $\\
 CS97   & $ 27    \pm 9    $   &  $<16             $  & $ 16.4    \pm 4.1  $ & $ 6.1   \pm 1.4  $  & $ <8.6            $   & $ 5.9   \pm 1.0 $    & $ 4.45   \pm  0.23$ & $ 2.11   \pm 0.26  $\\
 CS114  & $ <29            $   &  $<17             $  & $ 31.1    \pm 7.8  $ & $ 16.1  \pm 3.7  $  & $ 14.8   \pm 3.6  $   & $ 11.1  \pm 1.4 $    & $ 7.34   \pm  0.38$ & $ 3.02   \pm 0.36  $\\
 CS164  & $ <8             $   &  $<16             $  & $ 0.8     \pm 0.4  $ & $ 1.8   \pm 0.6  $  & $ <6.4            $   & $ 1.7   \pm 0.8 $    & $ 1.29   \pm  0.07$ & $ 0.81   \pm 0.16  $\\
 CS194  & $ 63    \pm 28   $   &  $<27             $  & $ 23.8    \pm 6.0  $ & $ 11.5  \pm 2.9  $  & $ <14.1           $   & $ 7.4   \pm 1.7 $    & $ 5.98   \pm  0.31$ & $ 3.10   \pm 0.37  $\\
 CS215  & $ <195^\star      $  & $<146^\star        $  & $ <102.4^\star      $ & $     \ldots     $  & $ <32.6^\star      $   & $ <51.0^\star    $    & $2.03    \pm  0.11$ & $ <13.29^\star     $\\
 CS241  & $ <13            $   &  $<24             $  & $ 8.1     \pm 2.1  $ & $ 1.2   \pm 0.3  $  & $ <7.3            $   & $ <1.8          $    & $ 1.08   \pm  0.06$ & $ 0.43   \pm 0.14  $\\
 CS265  & $ 161   \pm 41   $   &  $104  \pm 10     $  & $ 91.0    \pm 22.7 $ & $ 27.8  \pm 7.0  $  & $ 29.9   \pm 3.8  $   & $ 29.7  \pm 3.2 $    & $ 18.80  \pm  0.82$ & $ 12.10  \pm 1.22  $\\
 CS292  & $ 252   \pm 63   $   &  $159  \pm 8      $  & $ 104.2   \pm 26.1 $ & $ 60.7  \pm 15.2 $  & $ 38.9   \pm 4.3  $   & $ 37.3  \pm 3.9 $    & $ 21.99  \pm  0.80$ & $ 12.22  \pm 1.27  $\\
 CS415  & $ <7             $   &  $<17             $  & $ 8.3     \pm 2.1  $ & $     \ldots     $  & $ <5.5            $   & $ <7.1^\star     $    & $1.38   \pm  0.08 $ & $ 0.90   \pm 0.22  $\\
 CS520  & $ 37    \pm 10   $   &  $<17             $  & $ 18.8    \pm 4.7  $ & $     \ldots     $  & $ 7.0    \pm 4.1  $   & $ 5.0   \pm 0.9 $    & $ 4.19   \pm  0.22$ & $ 2.17   \pm 0.27  $\\
 CS538  & $ <9             $   &  $<16             $  & $ 8.2     \pm 2.1  $ & $     \ldots     $  & $ <5.4            $   & $ <3.9^\star     $    & $1.14   \pm  0.06 $ & $ <1.06^\star      $\\
 CS539  & $ 82    \pm 21   $   &  $57   \pm 6      $  & $ 45.0    \pm 11.3 $ & $     \ldots     $  & $ 18.3   \pm 4.3  $   & $ 14.2  \pm 1.6 $    & $ 9.82  \pm   0.50$ & $ 4.72   \pm 0.53  $\\
 CS574  & $ 55    \pm 12   $   &  $46   \pm 7      $  & $ 49.4    \pm 12.3 $ & $     \ldots     $  & $ 22.9   \pm 3.9  $   & $ 19.1  \pm 2.0 $    & $ 12.35 \pm   0.63$ & $ 8.24   \pm 0.84  $\\
 CS603  & $ 78    \pm 20   $   &  $58   \pm 6      $  & $ 52.5    \pm 13.1 $ & $     \ldots     $  & $ 28.5   \pm 5.0  $   & $ 18.6  \pm 2.0 $    & $ 11.90 \pm   0.61$ & $ 9.55   \pm 0.97  $\\
 CS618  & $ 441   \pm 111  $   &  $298  \pm 10     $  & $ 236.3   \pm 59.1 $ & $ 69.2  \pm 17.3 $  & $ 85.8   \pm 9.2  $   & $ 84.8  \pm 8.6 $    & $ 47.47 \pm   1.68$ & $ 25.90  \pm 2.65  $\\
 CS649  & $ <7             $   &  $<17             $  & $ 16.8    \pm 4.2  $ & $     \ldots     $  & $ <6.5            $   & $ 8.5   \pm 1.2 $    & $ 6.26  \pm   0.32$ & $ 3.88   \pm 0.43  $\\
 CS703  & $ 232   \pm 58   $   &  $149  \pm 7      $  & $ 128.5   \pm 32.1 $ & $ 43.5  \pm 10.9 $  & $      \ldots     $   & $ 42.1  \pm 4.3 $    & $ 22.93 \pm   1.18$ & $ 15.18  \pm 1.56  $\\
 CS713  & $ 69    \pm 18   $   &  $45   \pm 7      $  & $ 54.4    \pm 13.6 $ & $ 30.0  \pm 7.5  $  & $      \ldots     $   & $ 22.6  \pm 2.4 $    & $ 16.62 \pm   0.85$ & $ 11.42  \pm 1.18  $\\ \hline
 \end{tabular}
\tablefoot{Flux density upper limits resulting from confusion are marked by ($^\star$). Sources for which the flux density could not be measured at the respective frequency because of image artefacts are marked by ``af''. Sources that were outside the survey fields or not targeted by the observations at the respective frequency are represented by ellipsis dots~($\ldots$).}
\end{table*}
\begin{table*}
\ContinuedFloat
\caption{continued.}
\centering
 \begin{tabular}{c c c c c c c c c}
 \hline \hline
 IFRS  & $S_\mathrm{4.8\,GHz}$ & $S_\mathrm{5.5\,GHz}$ & $S_\mathrm{8.6\,GHz}$  & $S_\mathrm{9\,GHz}$ & $S_\mathrm{18\,GHz}$ & $S_\mathrm{20\,GHz}$ & $S_\mathrm{34\,GHz}$                 & $\alpha_{1.40}^{1.71}$ \\
 ID    & $\mathrm{[mJy]}$     & [mJy]                & [mJy]                 & [mJy]              & [mJy]               & [mJy]               & [mJy]                               &                       \\ \hline
 CS94  & $     \ldots     $   & $     \ldots      $  & $    \ldots     $     & $    \ldots     $  & $    \ldots     $   & $    \ldots     $   & $    \ldots      $                  & $ -0.79 \pm 0.17$     \\
 CS97  & $     \ldots     $   & $     \ldots      $  & $    \ldots     $     & $    \ldots     $  & $    \ldots     $   & $ <2.0          $   & $    \ldots      $                  & $ -0.91 \pm 0.10$     \\
 CS114 & $ 2.07  \pm 0.36 $   & $     \ldots      $  & $ 0.50 \pm 0.19 $     & $    \ldots     $  & $    \ldots     $   & $ <2.0          $   & $ 0.120\pm 0.036\tablefootmark{a} $ & $ -1.33 \pm 0.02$     \\
 CS164 & $ 0.76  \pm 0.19 $   & $     \ldots      $  & $ 0.21 \pm 0.11 $     & $    \ldots     $  & $    \ldots     $   & $ <2.0          $   & $    \ldots      $                  & $ -0.26 \pm 0.13$     \\
 CS194 & $ 1.69  \pm 0.23 $   & $     \ldots      $  & $ 1.19 \pm 0.25 $     & $    \ldots     $  & $    \ldots     $   & $ <2.0          $   & $ 0.205\pm 0.034 $                  & $ -1.01 \pm 0.05$     \\
 CS215 & $ 0.73  \pm 0.17 $   & $     \ldots      $  & $ 0.28 \pm 0.15 $     & $    \ldots     $  & $    \ldots     $   & $    \ldots     $   & $    \ldots      $                  & $ -0.71 \pm 0.04$     \\
 CS241 & $ <0.24          $   & $     \ldots      $  & $ <0.21         $     & $    \ldots     $  & $    \ldots     $   & $ <2.0          $   & $    \ldots      $                  & $ -1.05 \pm 0.19$     \\
 CS265 & $     \ldots     $   & $ 6.33  \pm  0.34 $  & $    \ldots     $     & $ 3.83 \pm 0.25 $  & $ 2.18 \pm 0.16 $   & $ 2.42 \pm 0.34 $   & $    \ldots      $                  & $ -0.71 \pm 0.01$     \\
 CS292 & $     \ldots     $   & $     \ldots      $  & $    \ldots     $     & $    \ldots     $  & $    \ldots     $   & $    \ldots     $   & $    \ldots      $                  & $ -0.79 \pm 0.02$     \\
 CS415 & $ <0.33          $   & $ 0.429 \pm  0.020$  & $ <0.27         $     & $    \ldots     $  & $    \ldots     $   & $ <2.0          $   & $    \ldots      $                  & $ -1.19 \pm 0.56$     \\
 CS520 & $     \ldots     $   & $ 1.292 \pm  0.013$  & $    \ldots     $     & $    \ldots     $  & $    \ldots     $   & $ <2.0          $   & $    \ldots      $                  & $ -1.02 \pm 0.08$     \\
 CS538 & $ <0.21          $   & $     \ldots      $  & $ <0.27         $     & $    \ldots     $  & $    \ldots     $   & $ <2.0          $   & $    \ldots      $                  & $ -1.19 \pm 0.08$     \\
 CS539 & $     \ldots     $   & $     \ldots      $  & $    \ldots     $     & $    \ldots     $  & $    \ldots     $   & $    \ldots     $   & $    \ldots      $                  & $ -0.89 \pm 0.04$     \\
 CS574 & $     \ldots     $   & $     \ldots      $  & $    \ldots     $     & $    \ldots     $  & $    \ldots     $   & $ <2.0          $   & $    \ldots      $                  & $ -0.81 \pm 0.01$     \\
 CS603 & $     \ldots     $   & $ 6.90  \pm  0.39 $  & $    \ldots     $     & $ 5.00 \pm 0.32 $  & $ 4.30 \pm 0.29 $   & $ 5.25 \pm 0.48 $   & $    \ldots      $                  & $ -0.65 \pm 0.01$     \\
 CS618 & $     \ldots     $   & $     \ldots      $  & $    \ldots     $     & $    \ldots     $  & $    \ldots     $   & $    \ldots     $   & $    \ldots      $                  & $ \ldots        $     \\
 CS649 & $     \ldots     $   & $     \ldots      $  & $    \ldots     $     & $    \ldots     $  & $    \ldots     $   & $ <2.0          $   & $    \ldots      $                  & $ -0.42 \pm 0.03$     \\
 CS703 & $ 8.60  \pm 0.82 $   & $     \ldots      $  & $ 4.22 \pm 0.37 $     & $    \ldots     $  & $    \ldots     $   & $    \ldots     $   & $ 1.150\pm 0.118 $                  & $ -0.97 \pm 0.01$     \\
 CS713 & $     \ldots     $   & $     \ldots      $  & $    \ldots     $     & $    \ldots     $  & $    \ldots     $   & $    \ldots     $   & $    \ldots      $                  & $ -0.61 \pm 0.03$     \\ \hline
 \end{tabular}
\tablefoot{\tablefoottext{a}{Detection at 33\,GHz.}}
\end{table*}

\end{document}